\documentclass[aps,preprint,nofootinbib,showpacs]{revtex4}
\usepackage[dvips]{graphicx,color}
\usepackage{amsmath}
\newcommand{\be}{\begin{equation}}
\newcommand{\ee}{\end{equation}}
\newcommand{\ba}{\begin{eqnarray}}
\newcommand{\beq}{\begin{equation}}
\newcommand{\eeq}{\end{equation}}
\newcommand{\ea}{\end{eqnarray}}

\newcommand{\eV}{\text{eV}}
\newcommand{\MeV}{\text{MeV}}
\newcommand{\GeV}{\text{GeV}}
\newcommand{\TeV}{\text{TeV}}
\newcommand{\fb}{\text{fb}}

\newcommand{\BR}{\text{BR}}

\newcommand{\mll}{{m_{\ell\ell^\prime}^{}}}
\newcommand{\hll}{{h_{\ell\ell^\prime}^{}}}
\newcommand{\qqHppHmm}{{q\overline{q}\to \gamma^*,Z^* \to H^{++}H^{--}}}

\newcommand{\qqHpmpmHmp}{{q^\prime\overline{q}\to W^* \to H^{\pm\pm}H^\mp}}

\newcommand{\ppHpmH}{{pp\to W^* \to H^\pm H^0}}
\newcommand{\ppHpH}{{pp\to W^* \to H^+H^0}}
\newcommand{\ppHmH}{{pp\to W^* \to H^-H^0}}

\newcommand{\ppHpmA}{{pp\to W^* \to H^\pm A^0}}
\newcommand{\ppHA}{{pp\to Z^* \to H^0 A^0}}
\newcommand{\Hpmlpmnu}{{H^\pm \to \ell^\pm \nu_{\ell^\prime}^{}}}

\def\beqa{\begin{eqnarray}}
\def\eeqa{\end{eqnarray}}

\def\bea{\begin{eqnarray}}
\def\eea{\end{eqnarray}}

\def\err#1#2{\lower2pt\hbox{ $\stackrel{\scriptstyle +#1}{\scriptstyle -#2}$}}
\def\ga{\mathrel{\raise.3ex\hbox{$>$\kern-.75em\lower1ex\hbox{$\sim$}}}}
\def\la{\mathrel{\raise.3ex\hbox{$<$\kern-.75em\lower1ex\hbox{$\sim$}}}}
\def\bmaT{\left(\begin{array}{ccc}}
\def\emaT{\end{array}\right)}
\def\bma{\left( \begin{array} }
\def\ema{\end{array} \right)}
\def\gsim{~{\rlap{\lower 3.5pt\hbox{$\mathchar\sim$}}\raise 1pt\hbox{$>$}}\,}
\def\lsim{~{\rlap{\lower 3.5pt\hbox{$\mathchar\sim$}}\raise 1pt\hbox{$<$}}\,}

\begin{document}

\preprint{
\vbox{%
\hbox{SHEP-11-32}
}}
\title{\boldmath Five-lepton and six-lepton signatures from production \\of 
neutral triplet scalars in the Higgs Triplet Model \unboldmath} 
\author{A.G. Akeroyd}
\email{a.g.akeroyd@soton.ac.uk}
\author{S. Moretti}
\email{S.Moretti@soton.ac.uk}
\affiliation{School of Physics and Astronomy, University of Southampton \\
Highfield, Southampton SO17 1BJ, United Kingdom}
\affiliation{Particle Physics Department, Rutherford Appleton Laboratory, 
Chilton, Didcot, Oxon OX11 0QX, United Kingdom}
\author{Hiroaki Sugiyama}
\email{hiroaki@fc.ritsumei.ac.jp}
\affiliation{Department of Physics,
Ritsumeikan University, Kusatsu, Shiga 525-8577, Japan}

\begin{abstract}

 The neutral triplet scalars~($H^0,A^0$)
in the Higgs Triplet Model~(HTM) are difficult to detect
at the LHC for the case of mass degeneracy
with the singly charged~($H^\pm$)
and doubly charged~($H^{\pm\pm}$) scalars. 
 A non-zero value of a specific quartic coupling
in the scalar potential
removes this degeneracy,
and positive values of this coupling
would give rise to the mass hierarchy 
$m_{A^0,H^0}^{} > m_{H^\pm}^{} > m_{H^{\pm\pm}}^{}$.
 In this scenario
the decays $H^0\to H^\pm W^*$ and $A^0\to H^\pm W^*$,
followed by $H^\pm\to H^{\pm\pm}W^*$,
can proceed with large branching ratios,
even for small mass splittings among the triplet scalars.
 Such a cascade process would lead to production of $H^{++}H^{--}$
accompanied by several virtual $W$ bosons,
and would provide a way of observing $H^0$ and $A^0$
with a signature which is different from
that of the conventional production process
for doubly charged scalars, $\qqHppHmm$.
 Assuming the decay $H^{\pm\pm}\to \ell^\pm{\ell^\prime}^\pm$,
we quantify the magnitude of
the five-lepton and six-lepton signals in these channels,
which have essentially negligible backgrounds at the LHC\@.

\end{abstract}
\maketitle


\section{Introduction} 
 The established evidence that
neutrinos oscillate and possess small masses~\cite{Fukuda:1998mi}
necessitates physics beyond the Standard Model~(SM),
which could  manifest itself at the CERN Large Hadron Collider~(LHC) 
and/or in low energy experiments
which search for lepton flavour violation~\cite{Kuno:1999jp}.
 Consequently,
models of neutrino mass generation which can be probed
at present and forthcoming experiments
are of great phenomenological interest.

 Neutrinos may obtain mass
via the vacuum expectation value~(vev)
of a neutral Higgs boson
in an isospin triplet representation%
~\cite{Konetschny:1977bn, Mohapatra:1979ia,
Magg:1980ut,Schechter:1980gr,Cheng:1980qt}.
 A particularly simple implementation
of this mechanism of neutrino mass generation
is the  ``Higgs Triplet Model''~(HTM)
in which the SM Lagrangian is augmented
solely by an $SU(2)_L$ triplet of scalar particles (denoted by $\Delta$)
with hypercharge $Y=2$%
~\cite{Konetschny:1977bn, Schechter:1980gr,Cheng:1980qt}.
 In the HTM,
the Majorana neutrino mass matrix $m_{\ell\ell^\prime}^{}$
($\ell,\ell^\prime=e,\mu,\tau$)
is given by the product of
a triplet Yukawa coupling matrix $\hll$
and a triplet vev ($v_\Delta$).
 Consequently,
this direct connection between $\hll$ and $m_{\ell\ell^\prime}^{}$
gives rise to phenomenological predictions
for processes which depend on $\hll$
because $\mll$ has been severely restricted  
by neutrino oscillation measurements%
~\cite{Fukuda:1998mi,solar,atm,acc,Apollonio:2002gd,:2008ee}.
 A distinctive signal of the HTM would be
the observation of a ``doubly charged Higgs boson'' ($H^{\pm\pm}$,
with two units of electric charge),
whose mass ($m_{H^{\pm\pm}}$) may be
of the order of the electroweak scale.
 Such particles could be produced with sizeable rates
at hadron colliders through the processes $\qqHppHmm$%
~\cite{Barger:1982cy, Gunion:1989in, Muhlleitner:2003me, Han:2007bk, Huitu:1996su}
and $\qqHpmpmHmp$~\cite{Barger:1982cy, Dion:1998pw, Akeroyd:2005gt}.

 The branching ratios~(BRs) for 
$H^{\pm\pm}\to \ell^\pm{\ell^\prime}^\pm$ depend on $\hll$
and are predicted in the HTM in terms of the parameters
of the neutrino mass matrix~\cite{Akeroyd:2005gt, Ma:2000wp, Chun:2003ej}.
 Detailed quantitative studies of
BR($H^{\pm\pm}\to \ell^\pm{\ell^\prime}^\pm$) in the HTM
have been performed in
Refs.~\cite{Garayoa:2007fw,Akeroyd:2007zv,Kadastik:2007yd,
Perez:2008ha,delAguila:2008cj,Akeroyd:2009hb} 
with particular emphasis  
given to their sensitivity to the Majorana phases and 
the absolute neutrino mass i.e.\ parameters which cannot be 
probed in neutrino oscillation experiments. 
 A study on the relation
between BR($H^{\pm\pm}\to \ell^\pm{\ell^\prime}^\pm$)
and the neutrinoless double beta decay
was performed in Ref.~\cite{Petcov:2009zr}.

 The first searches for $H^{\pm\pm}$ at a hadron collider 
were carried out at the Fermilab Tevatron,
assuming the production channel
$\qqHppHmm$ and the decay $H^{\pm\pm}\to \ell^\pm{\ell^\prime}^\pm$. 
 In the early searches~\cite{Acosta:2004uj,Aaltonen:2008ip}
the mass limits $m_{H^{\pm\pm}}> 110\to 150\,\GeV$ were derived,
with the strongest limits being for $\ell=e,\mu$~\cite{Acosta:2004uj}.
 The latest searches~\cite{Aaltonen:2011rta,Abazov:2011xx}
use $6\to 7\,\fb^{-1}$ of integrated luminosity,  
and the mass limits have improved
to $m_{H^{\pm\pm}}> 200\to 245\,\GeV$
for $\ell=e,\mu$~\cite{Aaltonen:2011rta}.
%
 Simulations of the detection prospects of $H^{\pm\pm}$ at the LHC 
with $\sqrt s=14\,\TeV$ previously focussed on
$\qqHppHmm$ only~\cite{Azuelos:2005uc},
but subsequent studies included
the mechanism $\qqHpmpmHmp$~\cite{Perez:2008ha, delAguila:2008cj, Akeroyd:2010ip}. 
 The first searches
for $H^{\pm\pm}$ at the LHC with $\sqrt s=7\,\TeV$
have recently been performed. 
 The CMS collaboration
(with $0.98\,\fb^{-1}$ of integrated luminosity)
has carried out separate searches
for $\qqHppHmm$ and $\qqHpmpmHmp$~\cite{CMS-search},
assuming the decay channels $H^{\pm\pm}\to \ell^\pm{\ell^\prime}^\pm$
and $\Hpmlpmnu$ with  $\ell=e,\mu,\tau$.
 The ATLAS collaboration has carried out three distinct searches
for the decay $H^{\pm\pm}\to \ell^\pm{\ell^\prime}^\pm$
(assuming production via $\qqHppHmm$ only) as follows:
 i) two (or more) leptons (for $\ell=\mu$ only),
using $1.6\,\fb^{-1}$ of integrated luminosity%
~\cite{Aad:2012cg};
 ii) three (or more) leptons ($\ell=e,\mu$),
using $1.02\,\fb^{-1}$ of integrated luminosity%
~\cite{ATLAS-search:3l};
 iii) four leptons ($\ell=e,\mu$),
using $1.02\,\fb^{-1}$ of integrated luminosity~\cite{ATLAS-search:4l}.
 The mass limits on  $m_{H^{\pm\pm}}$ from the LHC searches are
stronger than those from the Tevatron searches,
and are of the order of $m_{H^{\pm\pm}}> 300\,\GeV$ for $\ell=e,\mu$.
 If the decay channel $H^{\pm\pm}\to W^\pm W^\pm$ dominates
(which is the case for $v_\Delta > 10^{-4}\,\GeV$)
then the signature of $H^{\pm\pm}$ is different, 
for which there have been no direct searches.

 In addition to the above charged scalars
there are three electrically neutral Higgs scalars in the HTM:
 $h^0$ and $H^0$ are CP-even,
and $A^0$ is CP-odd.
 These scalar eigenstates are
mixtures of the doublet and triplet neutral fields,
but the mixing angle is very small
in most of the parameter space of the HTM
because of the hierarchy of the vevs, $v_\Delta \ll v$ 
(where $v=246\,\GeV$, the vev of the neutral doublet field).
 The phenomenology of the dominantly doublet scalar,
which we call $h^0$, is essentially the same as
that of the SM Higgs boson
(for which direct searches are ongoing),
apart from the following cases: 
 i) the scenario of charged scalar loops ($H^{\pm\pm}$ and $H^\pm$) 
substantially enhancing or suppressing the decay rate
for the decay $h^0\to \gamma\gamma$~\cite{Arhrib:2011vc},
which is possible if a scalar trilinear coupling
(such as $h^0 H^{++}H^{--}$) is sizeable;
 ii) the scenario of a large mixing angle in the CP-even sector,
which arises if $m_{h^0}\sim m_{H^0}$~\cite{Dey:2008jm, Akeroyd:2010je, Arhrib:2011uy};
 iii) the scenario of $m_{h^0}$ being heavy enough
to decay to two on-shell triplet scalars%
~\cite{Akeroyd:2011ir,Melfo:2011nx}
(e.g.\ $h^0\to H^{++}H^{--}$),
which would suppress BR$(h^0\to WW,ZZ)$
with respect to the SM prediction. 
 In the latest searches for the Higgs boson of the SM
by the CMS~\cite{CMS-combination}
and ATLAS collaborations~\cite{ATLAS-combination},
a slight excess  of events in the mass interval
of $125\,\GeV$ to $127\,\GeV$ has been reported.
 If this excess is substantiated with more data,
one interpretation would be $h^0$ in the HTM,
as emphasised in Ref.~\cite{Arhrib:2011vc}.

 So far there has been no search for
the neutral scalars $H^0$ and $A^0$, 
which can be produced at hadron colliders via the processes
$q^\prime\overline{q}\to W^* \to H^\pm H^0$, 
$q^\prime\overline{q}\to W^*\to H^\pm A^0$,
and $q\overline q\to Z^* \to H^0 A^0$~\cite{Akeroyd:2011zz,Melfo:2011nx,Aoki:2011pz}
(for studies of these processes in the Two Higgs Doublet Model,
see Ref.~\cite{Kanemura:2001hz}).
However, the detection prospects in these channels in the HTM are
not very promising in the 7 TeV run of the LHC
due to the moderate cross sections
and the fact that several of the expected decays channels
(e.g.\ $H^0,A^0\to \nu\nu,b\overline b$%
~\cite{Perez:2008ha})
suffer from large backgrounds.  
 In the HTM,
all the triplet scalars are approximately degenerate 
($m_{A^0,H^0}\sim  m_{H^\pm} \sim m_{H^{\pm\pm}}$)
if a specific quartic coupling in the scalar potential is zero
(i.e.\ $\lambda_4=0$).
 In general,
one expects $\lambda_4\ne 0$,
which would give rise to two distinct mass hierarchies:
 $m_{A^0,H^0}> m_{H^\pm} > m_{H^{\pm\pm}}$ for $\lambda_4> 0$ 
and $m_{A^0,H^0}< m_{H^\pm} < m_{H^{\pm\pm}}$ for  $\lambda_4 < 0$.
 The scenario of $m_{A^0,H^0}> m_{H^\pm} > m_{H^{\pm\pm}}$
enables the decay modes 
$H^0\to H^\pm W^*$ and $A^0\to H^\pm W^*$~\cite{Chun:2003ej} to proceed.
 We perform a numerical study of the magnitude
of BR($H^0\to H^\pm W^*$) and BR($A^0\to H^\pm W^*$) in the HTM,
and we show that
these three-body decay channels can have larger BRs
than the conventional two-body decay channels~\cite{Perez:2008ha}
of $H^0,A^0$ (for analogous results in certain Two Higgs Doublet Models
see Ref.~\cite{Akeroyd:1998dt}).
Such large values of BR($H^0\to H^\pm W^*$) and BR($A^0\to H^\pm W^*$)
could enhance the detection prospects for $H^0$ and $A^0$ at the LHC,
and this possibility 
was first mentioned (although not quantified) in Ref.~\cite{Akeroyd:2011zz}.

 The production processes
$q^\prime\overline{q}\to W^* \to H^\pm H^0$, 
$q^\prime\overline{q}\to W^*\to H^\pm A^0$,
and $q\overline q\to Z^* \to H^0 A^0$,
together with the above decays
(and subsequently the decay $H^\pm\to H^{\pm\pm} W^*$,
which can also have a large BR~\cite{Akeroyd:2011zz}) 
would lead to production of $H^{++}H^{--}$ accompanied
by several $W^*$ bosons.
 This additional production of $H^{++}H^{--}$ pairs
from the decays of $H^\pm$, $H^0$ and $A^0$ would increase
the sensitivity to $m_{H^{\pm\pm}}$ in direct searches for $H^{\pm\pm}$ at the LHC\@.
 A simulation which includes all of the above production mechanisms
for $H^{\pm\pm}$, $H^\pm$, $A^0$ and $H^0$
has been performed in Ref.~\cite{Melfo:2011nx},
using the same selection cuts
as for the four-lepton search for $\qqHppHmm$
in Ref.~\cite{CMS-search},
and it was shown that
the mass limits on $m_{H^{\pm\pm}}$ could be increased
by as much as $50\,\GeV$.
Importantly, the additional $W^*$ from the cascade decays
would provide a way of observing 
$H^0$ and $A^0$ with a signature which is different from
that of the conventional production process
for doubly charged scalars ($\qqHppHmm$),
and this issue was not the focus of the study in Ref.~\cite{Melfo:2011nx},
which was concerned with the discovery of $H^{++}H^{--}$
in the four-lepton channel only.
In this paper we quantify the magnitude of the cross sections
for the production of $H^0$ and $A^0$ with subsequent decay
to $H^{++}H^{--}$ and several $W^*$,
and we advocate separate simulations of these signatures
as a way of discovering $H^0$ and $A^0$
(assuming initial discovery of  $H^{++}H^{--}$
in the channels with four or fewer leptons).
 Of particular interest are the five-lepton (i.e.\ $W^*\to \ell\nu$) 
and six-lepton channels (i.e.\ $W^*W^*\to \ell\nu\ell\nu$), 
for which the background is known to
very small or negligible~\cite{delAguila:2008cj}.
 Importantly, the six-lepton channel
can only arise from production mechanisms
which involve $H^0$ and $A^0$.

 Our work is organised as follows.
 In section~II we briefly describe
the theoretical structure of the HTM\@. 
 In section~III
the magnitude of BR($H^0\to H^\pm W^*$) and BR($A^0\to H^\pm W^*$)
is studied as a function of the parameters of the scalar potential.
 Section~IV contains our numerical analysis of the magnitude
of the cross section for
$H^{++}H^{--}$ plus several $W^*$,
 with emphasis given to the five-lepton and six-lepton signatures.
 Conclusions are given in section~V\@.

\section{The Higgs Triplet Model}

 In the HTM~\cite{Konetschny:1977bn,Schechter:1980gr,Cheng:1980qt}
a $Y=2$ complex $SU(2)_L$ isospin triplet of scalar fields,
${\bf T}=( T_1, T_2, T_3 )$, is added to the SM Lagrangian. 
 Such a model can provide Majorana masses for the observed neutrinos 
without the introduction of $SU(2)_L$ singlet neutrinos
via the gauge invariant Yukawa interaction:
\begin{equation}
{\cal L}=\hll L_\ell^TCi\sigma_2\Delta L_{\ell^\prime}+\text{h.c.}
\label{trip_yuk}
\end{equation}
 Here $\hll (\ell,\ell^\prime=e,\mu,\tau)$ is a complex
and symmetric coupling,
$C$ is the Dirac charge conjugation operator,
$\sigma_i (i=1\text{-}3)$ are the Pauli matrices,
$L_\ell=(\nu_{\ell L}, \ell_L)^T$ is a left-handed lepton doublet,
and $\Delta$  is a $2\times 2$ representation
of the $Y=2$ complex triplet fields:
\begin{equation}
\Delta
= {\bf T}\cdot\frac{\sigma}{2}
= T_1 \frac{\sigma_1}{2} + T_2 \frac{\sigma_2}{2} + T_3 \frac{\sigma_3}{2}
=\bma{cc}
\Delta^+/\sqrt{2}  & \Delta^{++} \\
\Delta^0       & -\Delta^+/\sqrt{2}
\ema ,
\end{equation}
where
$T_1 = (\Delta^{++} + \Delta^0)$,
$T_2 = i(\Delta^{++} - \Delta^0)$,
and $T_3 = \sqrt{2}\,\Delta^+$.
 A non-zero triplet vev $\langle\Delta^0\rangle$ 
gives rise to the following mass matrix for neutrinos:
\begin{equation}
\mll = 2\hll \langle\Delta^0\rangle = \sqrt{2}\hll v_{\Delta} .
\label{nu_mass}
\end{equation}
 The necessary non-zero $v_{\Delta}$ arises from the minimisation
of the most general $SU(2)_L\otimes U(1)_Y$ invariant Higgs potential%
~\cite{Cheng:1980qt,Gelmini:1980re},
which is written%
\footnote{
 One may rewrite the potential in eq.~(\ref{Potential})
by using
$2 \text{Det}(\Delta^\dagger \Delta)
= \text{Tr}[\Delta^\dagger \Delta^\dagger \Delta \Delta]
= [\text{Tr}(\Delta^\dagger \Delta)]^2
- \text{Tr}[(\Delta^\dagger \Delta)^2]$,
$(H^\dagger \sigma_i H) \text{Tr}(\Delta^\dagger \sigma_i \Delta)
= 2 H^\dagger \Delta \Delta^\dagger H
-(H^\dagger H) \text{Tr}(\Delta^\dagger \Delta)$,
and
$H^\dagger \Delta^\dagger \Delta H
= (H^\dagger H) \text{Tr}(\Delta^\dagger \Delta)
- H^\dagger \Delta \Delta^\dagger H$.
}
as follows~\cite{Ma:2000wp, Chun:2003ej}
(with $H=(\phi^+,\phi^0)^T$):

\begin{eqnarray}
V(H,\Delta) & = & 
- m_H^2 \ H^\dagger H \ + \ \lambda (H^\dagger H)^2 \ 
+ \ M_{\Delta}^2 \ {\rm Tr} \Delta^\dagger \Delta\ 
+ \ \left( \mu \ H^T \ i \sigma_2 \ \Delta^\dagger H \ + \ {\rm h.c.}\right) \ 
\nonumber \\
&& + \ \lambda_1 \ (H^\dagger H) {\rm Tr} \Delta^\dagger \Delta \ 
+ \ \lambda_2 \ \left( {\rm Tr} \Delta^\dagger \Delta \right)^2 \ 
+ \ \lambda_3 \ {\rm Tr} \left( \Delta^\dagger \Delta \right)^2 \ 
+ \ \lambda_4 \ H^\dagger \Delta \Delta^\dagger H. 
\label{Potential}
\end{eqnarray}

Here $m_H^2<0$
in order to ensure non-zero $\langle\phi^0\rangle=v/\sqrt 2$
which spontaneously breaks
$SU(2)_L \otimes U(1)_Y$ to $U(1)_Q$
while $M^2_\Delta > 0$.
 In the model of Gelmini-Roncadelli~\cite{Gelmini:1980re} 
the term $\mu(H^Ti\sigma_2\Delta^\dagger H)$ is absent,
and a non-zero $v_\Delta$
is obtained only by a spontaneous violation
of lepton number for $M^2_\Delta<0$.
 The resulting Higgs spectrum contains
a massless triplet scalar (majoron, $J$)
and another light scalar ($H^0$),
with the coupling $ZH^0J$ being unsuppressed.
 Pair production via $e^+e^-\to Z^*\to H^0J$ would give
a large  contribution to the invisible width of the $Z$,
and thus this model was excluded
at the CERN Large Electron Positron Collider~(LEP). 
 The inclusion of the term
$\mu(H^Ti\sigma_2\Delta^\dagger H$)~\cite{Cheng:1980qt}
explicitly breaks lepton number $L\#$
when $\Delta$ is assigned $L\#=-2$.
 Then,
the majoron is eliminated from the model,
and a non-zero $v_\Delta$ can be obtained even for $M_\Delta^2 > 0$.
 Thus the scalar potential in eq.~(\ref{Potential})
together with the triplet Yukawa interaction of eq.~(\ref{trip_yuk})
lead to a phenomenologically viable model
of neutrino mass generation.
 For small $v_\Delta/v$,
the expression for $v_\Delta$
resulting from the minimisation of $V$ is:
\begin{equation}
v_\Delta
\simeq
 \frac{ \mu v^2 }
      { \sqrt{2} ( M^2_\Delta + v^2 (\lambda_1+\lambda_4)/2 ) }\,. 
\label{tripletvev}
\end{equation}

 For $M_\Delta \gg v$
one has $v_\Delta \simeq \mu v^2/(\sqrt{2}\,M^2_\Delta)$,
which would naturally lead to a small $v_\Delta$
even for $\mu$ of the order of the electroweak scale
(and is sometimes called the ``Type II seesaw mechanism'').
 However, in this scenario
the triplet scalars would be too heavy to be observed at the LHC\@. 
 In recent years
there has been much interest
in the case of light triplet scalars ($M_\Delta\approx v$) 
within the discovery reach of the LHC,
for which eq.~(\ref{tripletvev}) leads to $v_\Delta\approx \mu$,
and this is the scenario we will focus on.
 The case of $v_\Delta < 0.1\,\MeV$
is phenomenologically attractive
(and it is assumed in the ongoing searches at the LHC) 
because the BRs of the triplet scalars
to leptonic final states
(e.g.\ $H^{\pm\pm}\to \ell^\pm{\ell^\prime}^\pm$) would be $\sim 100\%$. 
Due to the condition $v_\Delta\approx \mu$ for light triplet scalars,
$\mu$ must also be small (compared to the electroweak scale)
for the scenario of $v_\Delta < 0.1\,\MeV$. 
 Since $\mu$ is the only source of 
explicit lepton number violation in the HTM,
its numerical value would be naturally small 
according to the 't~Hooft criteria.
 Moreover,
if one requires
that the triplet Yukawa couplings $h_{ij}$ are
greater in magnitude than the smallest Yukawa coupling in the SM
(i.e.\ the electron Yukawa coupling, $y_e\sim 10^{-6}$)
then from eq.~(\ref{nu_mass})
one has $v_\Delta < 0.1\,\MeV$,
and thus the decays of  the triplet scalars to leptonic final states
have BRs which sum to $\sim 100\%$.
 In extensions of the HTM
the term $\mu(H^Ti\sigma_2\Delta^\dagger H$) may arise in various ways:
 i) it can be generated at tree level
via the vev of a Higgs singlet field~\cite{Schechter:1981cv}; 
 ii) it can arise at higher orders
in perturbation theory~\cite{Chun:2003ej};
 iii) it can originate
in the context of extra dimensions~\cite{Ma:2000wp}.
 
 An upper limit on $v_\Delta$ can be obtained
from considering its effect on the parameter
$\rho (=M^2_W/M_Z^2\cos^2\theta_W)$. 
 In the SM $\rho=1$ at tree-level,
while in the HTM one has (where $x=v_\Delta/v$):
\begin{equation}
\rho
\equiv
 1 + \delta\rho = {1+2x^2 \over 1+4x^2} .
\label{deltarho}
\end{equation}
 The measurement $\rho\approx 1$ leads to the bound
$v_\Delta/v\lsim 0.03$, or  $v_\Delta\lsim 8\,\GeV$.
 Production mechanisms which depend on $v_\Delta$ 
(i.e.\ $pp\to W^{\pm *}\to W^\mp H^{\pm\pm}$
and fusion via 
$W^{\pm *} W^{\pm *} \to H^{\pm\pm}$%
~\cite{Huitu:1996su,Gunion:1989ci,Vega:1989tt})
have much smaller cross sections than the processes 
$\qqHppHmm$ and $\qqHpmpmHmp$
at the energies of the Fermilab Tevatron,
but such mechanisms could be the dominant source of 
$H^{\pm\pm}$ at the LHC
if $v_{\Delta}={\cal O}(1)\,\GeV$ and $m_{H^{\pm\pm}}> 500\,\GeV$.
 However,
we note that for $v_{\Delta}={\cal O}(1)\,\GeV$
the decay channel of $H^{\pm\pm}$ to leptonic final states
would be negligible,
and thus the production mechanisms which depend on $v_\Delta$
can be neglected in searches for $H^{\pm\pm}\to \ell^\pm{\ell^\prime}^\pm$. 
 At the 1-loop level,
$v_\Delta$ must be renormalised
and explicit analyses lead to bounds on its magnitude
similar to the above bound
from the tree-level analysis, e.g.\ see Ref.~\cite{Blank:1997qa}.

 The scalar eigenstates in the HTM are as follows:
 i) the charged scalars $H^{\pm\pm}$ and $H^\pm$;
 ii) the CP-even neutral scalars $h^0$ and $H^0$;
 iii) a CP-odd neutral scalar $A^0$.
The squared mass of $H^\pm$ is given by
$m_{H^\pm}^2 \simeq M_\Delta^2 + (2\lambda_1 + \lambda_4)v^2/4$.
The squared masses of $H^{\pm\pm}$, $H^0$, and $A^0$
are approximately given as follows, with a mass splitting $\lambda_4 v^2/4$~\cite{Ma:2000wp,Chun:2003ej,Akeroyd:2010je}:
\begin{eqnarray}
m_{H^{\pm\pm}}^2 + \frac{\lambda_4}{4} v^2
\ \simeq \
 m_{H^\pm}^2
\ \simeq \
 m_{H^0,A^0}^2 - \frac{\lambda_4}{4} v^2 .
\end{eqnarray}
 Note that $\lambda_1$ is a free parameter
even if the scalar masses are fixed
because it appears in the masses
as a combination $M_\Delta^2 + \lambda_1 v^2/2$.
 The degeneracy $m_{H^0} \simeq m_{A^0}$
can be understood by the fact that
the Higgs potential is invariant
under a global $U(1)$ symmetry for $\Delta$
(i.e.\ $L\#$ conservation)
when one neglects the trilinear term proportional to $\mu$.
When one also neglects $\lambda_4$,
the global $U(1)$ is increased to a global $SU(2)$ symmetry for $\Delta$,
which makes the masses of the triplet scalars degenerate.

 The mass hierarchy
$m_{H^{\pm\pm}}^{} < m_{H^\pm}^{} <  m_{H^0,A^0}^{}$
is obtained for $\lambda_4 > 0$,
and the opposite hierarchy
$m_{H^{\pm\pm}}^{} > m_{H^\pm}^{} > m_{H^0,A^0}^{}$
is obtained for $\lambda_4 < 0$.
 In general,
one would not expect degenerate masses
for $H^{\pm\pm},H^\pm,H^0,A^0$,
but instead one of the above two mass hierarchies.
 The sign of $\lambda_4$ is not fixed
by theoretical requirements of vacuum stability
of the scalar potential~\cite{Arhrib:2011uy},
although $|\lambda_4|< 2 m_{H^\pm}^2/v^2$ is necessary
to ensure that $m_{H^{\pm\pm}}^2$ and $m_{H^0,A^0}^2$ are positive.

 The doubly charged scalar $H^{\pm\pm}$ is entirely composed
of the triplet scalar field $\Delta^{\pm\pm}$, 
while the remaining eigenstates are in general mixtures
of the doublet and triplet fields.
 The mixing angles
$\theta_0$ (for $\text{Re}(\Delta^0)$ and $\text{Re}(\phi^0)$),
$\theta_A$ (for $\text{Im}(\Delta^0)$ and $\text{Im}(\phi^0)$),
and $\theta_+$ (for $\Delta^\pm$ and $\phi^\pm$)
are obtained as follows:
\begin{eqnarray}
\sin\theta_0
&\simeq&
 s_0^\prime \frac{v_\Delta}{v} , \quad
%
s_0^\prime
\equiv
 \frac{4M_\Delta^2}
      {2M_\Delta^2 + (-4\lambda + \lambda_1 + \lambda_4) v^2}
\simeq
 \frac{ 2 m_{H^{\pm\pm}}^2 - \lambda_1 v^2 }
      { m_{H^0}^2 - m_h^2 } ,
\\
%
%
\sin\theta_A
&=&
 2 \frac{v_\Delta}{v} ,
\\
%
%
\sin\theta_+
&=&
 \sqrt{2} \frac{v_\Delta}{v} .
\end{eqnarray}
 These mixings are small {\it even if} $v_\Delta$
assumes its largest value of a few GeV\@.%
\footnote{
 A large mixing angle is possible in the CP-even sector
provided that $m_{h^0}\simeq m_{H^0}$~\cite{Akeroyd:2010je,Dey:2008jm}.
}
 Therefore
$H^\pm,H^0,A^0$ are predominantly composed of the triplet fields,
while $h^0$ is predominantly composed of the doublet field
and plays the role of the SM Higgs boson.

\section{The phenomenology of the neutral scalars, $H^0$ and $A^0$, of the HTM
}

 The phenomenology of the neutral triplet scalars
($H^0$ and $A^0$) of the HTM
has received much less attention
than the phenomenology of the charged scalars
($H^{\pm\pm}$ and $H^\pm$).
 A comprehensive study of the BRs of $H^0$ and $A^0$
was performed in Ref.~\cite{Perez:2008ha}
for the case of all the triplet scalars
being degenerate (i.e.\ $\lambda_4=0$).
 For $\lambda_4>0$
the decays $H^0,A^0\to H^\pm W^*$ would be possible, 
and we will quantify the magnitude of their BRs in this section.

\subsection{Degenerate triplet scalars for $v_\Delta<10^{-4}\,\GeV$}
 For the case of degeneracy of the triplet scalars, 
the partial widths of the available decay modes for $H^0$ and $A^0$
are proportional to $v^2_{\Delta}$ or $1/{v_\Delta^2}$
(the same is true for the decay widths of $H^{\pm\pm}$ and $H^\pm$).
 Therefore
$v_\Delta$ is a crucial parameter
which determines the branching ratios of the triplet scalars.
 The only channel which is proportional to $1/{v_\Delta^2}$
is  $H^0,A^0 \to \nu\nu + \overline{\nu}\,\overline{\nu}$:
\begin{eqnarray}
\Gamma(H^0,A^0\to \nu_L\nu_L)
\simeq
 \Gamma(H^0,A^0\to \overline{\nu_L}\,\overline{\nu_L})
\simeq
 \frac{ m_{H^0,A^0} \sum_i m_i^2 }{ 32 \pi v_\Delta^2 } ,
\end{eqnarray}
where $m_i$~($i=1\text{-}3$) are neutrino masses.
 For $v_\Delta<10^{-4}\,\GeV$
the scalar eigenstates are essentially composed of the triplet fields,
and one has
BR($H^0,A^0\to \nu\nu+\overline{\nu}\,\overline{\nu})\sim 100\%$.
 The lowest value of $v_\Delta$ we consider is $100\,\eV$,
which is compatible with the existing bounds on $v_\Delta$
from lepton flavour violating processes
such as $\mu\to e\gamma$, $\mu\to \bar{e}ee$
and $\tau\to \bar{\ell}\ell^\prime \ell^{\prime\prime}$%
~\cite{Chun:2003ej,Kakizaki:2003jk,Akeroyd:2009nu,Fukuyama:2009xk}

 This scenario is experimentally very challenging at the LHC,
because production via $q\overline q\to Z^\ast \to H^0A^0$
would have an invisible signature. 
 The production channel $q'\overline q\to H^0W$,
although giving a visible signature, 
would have a negligible production rate
because the $H^0WW$ coupling is very suppressed
for $v_\Delta<10^{-4}\,\GeV$.
 Moreover,
$t$-channel diagrams which depend on the coupling $H^0q\overline q$
are suppressed by this small Yukawa coupling,
as well as other factors
such as small Cabbibo-Kobayashi-Maskawa matrix elements
and parton distribution functions for third generation quarks.  
 Similar comments apply to $q'\overline q\to A^0W$,
for which the coupling $A^0WW$ is absent at the tree level.
 The mechanisms
$q'\overline q\to W^\ast \to H^\pm H^0$
and  $q'\overline q\to W^\ast \to H^\pm A^0$,
although not being suppressed by a small coupling,
would give rise to a signature of 
$\ell^\pm$ and missing energy,
which suffers from large backgrounds.
 Therefore the neutral scalars which originate from $\Delta^0$,
with the vev of the CP-even neutral field
providing the mass for the neutrinos, 
would (most likely) be unobservable at the LHC
for $v_\Delta<10^{-4}\,\GeV$.

At an $e^+e^-$ collider
we note that it would be possible to observe
the process $e^+e^-\to H^0A^0\to 4\nu$
provided that a photon is radiated from the lepton beams.
 Using LEP data for ``a photon plus missing energy'' events,
the bound $m_{A^0}+m_{H^0}> 110\,\GeV$ was derived in Ref.~\cite{Datta:1999nc}.

 In contrast to the above,
the scenario of $v_\Delta<10^{-4}\,\GeV$ is the most favourable case
for the detection of the charged scalars at the LHC,
since it gives rise to 
BR$(H^{\pm\pm}\to \ell^\pm\ell^\pm)\sim 100\%$
and BR$(H^{\pm}\to \ell^\pm\nu_\ell)\sim 100\%$, 
for which searches are ongoing at the LHC\@.
 Hence for $v_\Delta < 10^{-4}\,\GeV$,
and assuming degeneracy of the triplet scalars ($\lambda_4=0$),
it is conceivable that
the LHC discovers $H^{\pm\pm}$ and $H^\pm$
but cannot observe $H^0$ and $A^0$.
 Below
we will show that
the potential dominance of the decays 
$H^0,A^0\to H^\pm W^*$ for the case of $\lambda_4 >0$
would give rise to a distinctive signature for $H^0$ and $A^0$,
which would enable their detection for the case
of $v_\Delta<10^{-4}\,\GeV$.

\subsection{Degenerate triplet scalars for $v_\Delta>10^{-4}\,\GeV$}
 For the converse case of $v_\Delta > 10^{-4}\,\GeV$
the decay channels $H^0,A^0\to \nu\nu + \overline{\nu}\,\overline{\nu}$
are negligible,
and both $H^0$ and $A^0$ have visible decays,
some of which have contributions
from both the doublet and triplet components.
 The decay channels whose partial
widths are proportional to $v_\Delta^2$
(and so dominate for $v_\Delta > 10^{-4}\,\GeV$)
are
$H^0 \to ZZ$, $W^+ W^-$, $h^0h^0$, $q\bar{q}$,
and $A^0 \to h^0Z$, $q\bar{q}$.

 The partial decay width of $H^0 \to ZZ$
is expressed as
\begin{eqnarray}
\Gamma(H^0 \to ZZ)
&=&
 \frac{ g_{HZZ}^2 v^2 m_{H^0}^3 }
      { 32 \pi m_Z^4 }
 \left(
  1
  - \frac{4 m_Z^2 }{ m_{H^0}^2 }
  + \frac{ 12 m_Z^4 }{ m_{H^0}^4 }
 \right)
 \left(
  1 - \frac{ 4 m_Z^2 }{ m_{H^0}^2 }
 \right)^{1/2} ,
\label{eq:Gam_Hp_WZ}\\
%
g_{HZZ}^{}
&\equiv&
 \frac{ g_2^2 }{ 4 c_w^2 }
 ( 4 v_\Delta \cos\theta_0 - v \sin\theta_0 ) .
\end{eqnarray}
 In the coupling $g_{HZZ}^{}$
(which appears in the Lagrangian as the term $g_{HZZ}^{}\,v\,H^0 Z^\mu Z_\mu$),
there is a contribution
from the doublet scalar component $\sim v \sin\theta_0$
and from the triplet scalar component
$\sim v_\Delta^{} \cos\theta_0$.
 The coupling $g_{HZZ}^{}$
can be rewritten as
$g_{HZZ}^{} \simeq (4-s_0^\prime) g_2^2 v_\Delta / (4 c_w^2)$.
 For $M_{\Delta}\gg v$
one has $\sin\theta_0 \simeq 2v_\Delta/v$.
 This value of $\sin\theta_0$ was used in the expressions for
the partial widths of the decay channels of $A^0$ and $H^0$
in Ref.~\cite{Perez:2008ha}.
 Although we do not assume $M_{\Delta}\gg v$
in our numerical analysis,
we take for simplicity $\sin\theta_0 = 2v_\Delta/v$,
which is possible with
$\lambda_1 \simeq 2(m_{h^0}^2 - m_{H^0}^2 + m_{H^{\pm\pm}}^2)/v^2$.
 For this value of $\sin\theta_0$
the two terms in $g_{HZZ}^{}$
are of a similar magnitude. Importantly,
the coupling $H^0WW$ vanishes at leading order
for $\sin\theta_0 \simeq 2v_\Delta/v$
due to an almost exact cancellation
between the doublet and triplet contributions,
$(v \sin\theta_0 - 2 v_\Delta \cos\theta_0)\simeq 0$.
 Consequently,
we will neglect the decay channel $H^0\to WW$,
as done in Ref.~\cite{Perez:2008ha}.
 If $\lambda_1 \simeq 2 (m_{H^{\pm\pm}}^2 - 2m_{H^0}^2 + 2m_{h^0}^2)/v^2$
then $s_0^\prime \simeq 4$ is possible, which leads to the opposite scenario of
a complete cancellation in the coupling which mediates $H^0\to ZZ$, and no such cancellation in the
coupling which mediates $H^0\to WW$.
 Note that
the decay channels $A^0\to ZZ$ and $A^0\to WW$ are absent
due to the CP-odd nature of $A^0$.

 If $m_{H^0}> 2m_{h^0}$
then the decay channel $H^0\to h^0h^0$ is open
with the following partial width:
\begin{eqnarray}
\Gamma(H^0 \to h^0h^0)
=
 \frac{\lambda^2_{Hhh} v^2}{8\pi m_{H^0}^{} }
 \left( 1 - \frac{ 4m_{h^0}^2 }{ m_{H^0}^2 } \right)^{1/2} .
\label{eq:Gam_Hp_Wh}
\end{eqnarray}
 The full expression for
$\lambda_{Hhh}$ (which appears in the Lagrangian as the term $\lambda_{Hhh} v H^0 h^0 h^0$) is complicated~\cite{Aoki:2011pz}
but the approximated form is as follows:
\begin{eqnarray}
\lambda_{Hhh}^{}
&\simeq&
 -
 \left\{
  M_\Delta^2
  + \frac{1}{\,4\,}
    \left(
     3\lambda - 4\lambda_1 - 4\lambda_4
    \right) s_0^\prime v^2
 \right\}
 \frac{v_\Delta}{v^3}
\simeq
 - \frac{1}{\,2\,} s_0^\prime
 ( m_{H^0}^2 - 2 m_{h^0}^2 )
 \frac{v_\Delta}{v^3} .
\end{eqnarray}

 The decays $H^0\to q\overline{q}$ proceed
via the doublet component only.
 Their decay rates are obtained from the expressions
for the analogous decays for the SM Higgs boson,
with multiplication by $\sin^2\theta_0$.
 Using $\sin\theta_0 \simeq s_0^\prime v_\Delta/v$
and neglecting QCD corrections
one obtains the following expressions for the partial widths:
\begin{eqnarray}
\Gamma(H^0 \to q\overline{q})
\simeq
 \frac{3 (s_0^\prime)^2 v_\Delta^2 G_F^2 m^2_q m_{H^0}^{}}{ 4 \pi }
 \left( 1 - \frac{ 4m^2_q }{ m_{H^0}^2 } \right)^{3/2} ,
\label{eq:Gam_H_qq}
\end{eqnarray}
where we use $s_0^\prime = 2$.
 If we take
$\lambda_1 \simeq 2 m_{H^{\pm\pm}}^2/v^2$,
we have $s_0^\prime \simeq 0$
($\sin\theta_0 = {\mathcal O}(v_\Delta^3/v^3)$),
and then $H^0 \to h^0 h^0$ and $q\bar{q}$ become negligible
while $H^0 \to ZZ$ and $WW$ are present.

 On the other hand,
the decay channels of $A^0$ into quarks
are as follows: 
\begin{eqnarray}
\Gamma(A^0 \to q\overline{q})
&\simeq&
 \frac{3 v_\Delta^2 G_F^2 m^2_q m_{A^0}}{ \pi }
 \left( 1 - \frac{ 4m^2_q }{ m_{A^0}^2 } \right)^{1/2} .
\label{eq:Gam_H_qq}
\end{eqnarray}
%
%
 If $m_{A^0}> m_{h^0}+m_Z$
then the decay channel $A^0\to h^0 Z$ is open.
\begin{eqnarray}
\Gamma(A^0 \to h^0 Z)
&=&
 \frac{ g_{AhZ}^2 m_{A^0}^3 }
      { 16 \pi m_Z^2 }
 \left(
  1 - \frac{ (m_{h^0} + m_Z)^2 }{ m_{A^0}^2 }
 \right)^{3/2}
 \left(
  1 - \frac{ (m_{h^0} - m_Z)^2 }{ m_{A^0}^2 }
 \right)^{3/2} ,\\
%
g_{AhZ}^{}
&\equiv&
 - \frac{g_2}{2c_w}
 \left(
  2 \sin\theta_0 \cos\theta_A
  - \cos\theta_0 \sin\theta_A
 \right)
\simeq
 - \frac{g_2}{c_w}
   \left(
    s_0^\prime - 1
   \right)
   \frac{v_\Delta}{v} .
\end{eqnarray}

 For the CP-even scalar,
the decay mode $H^0\to ZZ$ is the dominant channel
for $v_\Delta > 10^{-4}\,\GeV$ and $m_{H^0}>2m_Z$,
unless $H^0\to h^0h^0$ is open.
 For the CP-odd scalar
the decay mode $A^0\to h^0Z$ is the dominant channel,
if open kinematically.
 Otherwise,
$A^0\to b\overline{b}$ dominates for $m_{A^0}< 2m_t$
and $A^0\to t\overline{t}$ dominates $m_{A^0}> 2m_t$.

 From the production mechanism $q\overline{q}\to Z^\ast \to H^0A^0$,
the most promising signature would be $ZZZh^0$.
 From the mechanism $q'\overline q\to W^\ast \to H^\pm H^0$
one could have the signature of $WZZZ$,
and from $q'\overline q\to W^\ast \to H^\pm A^0$
the signature of $WZZh^0$.
 Detailed simulations would be needed to address
the feasibility of observing these signatures,
given the moderate cross sections,
although their detection prospects should be better than for the case
of $v_\Delta<10^{-4}\,\GeV$
(for which the decays
$H^0,A^0\to \nu\nu+\overline{\nu}\,\overline{\nu}$ dominate).
 We now give a rough quantitative estimate
of the discovery prospects of $A^0$ and $H^0$ 
for $v_\Delta > 10^{-4}\,\GeV$ at the LHC with $\sqrt s=7\,\TeV$.
 Importantly,
for $v_\Delta > 10^{-4}\,\GeV$
the decay $H^{\pm\pm}\to W^\pm W^\pm$ dominates and
there have been no direct searches for this channel
(for parton-level studies of detection prospects in this channel
see Refs.~\cite{Han:2007bk, Perez:2008ha}).
 Therefore
the masses of the degenerate triplet scalars could be
much lower than $300\,\GeV$.
 For $m_{H^0}^{}=150\,\GeV$ and $200\,\GeV$
the numerical values of $\sigma(q\overline q\to Z^\ast \to H^0A^0)$,
without the QCD $K$ factor, 
are $54\,\fb$ and $18\,\fb$ respectively.
 For the decay channel $H^0\to ZZ$ 
(which can have the dominant BR for $m_{H^0}^{}> 2m_Z^{}$)
followed by the decays $ZZ\to \ell^+\ell^- \nu\nu$ $(\ell=e,\mu$) 
the cross section would be of the order of $0.2\,\fb$ for
$m_{H^0}^{}=200\,\GeV$.
 Although this signature has a low background at the LHC
(and is similar to the signature
which arises from the production of the SM Higgs boson
in the channel $gg\to (h^0)^\ast \to ZZ$)
there would be at most 4 events before selection cuts
with $20\,\fb^{-1}$ of integrated luminosity.
 Therefore we conclude that
the discovery of $H^0$ and $A^0$ would be marginal at $\sqrt s=7\,\TeV$
for the case of degeneracy of the triplet scalars.
 Prospects are considerably better for $\sqrt s=14\,\TeV$
and larger ($> 100\,\fb^{-1}$) integrated luminosities,
with $\sigma(q\overline q\to Z^\ast\to H^0A^0)\simeq 72\,\fb$
for $m_{H^0}^{}=200\,\GeV$.

 Recently
the scenario of $m_{A^0,H^0}~{} < m_{H^\pm}^{} < m_{H^{\pm\pm}}^{}$
(for $\lambda_4 < 0$)
has been studied in Ref.~\cite{Aoki:2011pz},
taking $m_{H^0,A^0}^{}=119\,\GeV$ for which
$H^0/A^0\to b\overline b$  are the dominant decays.
 The cascade decay $H^{\pm\pm}\to H^\pm W^*$
followed by $H^\pm\to A^0/H^0 W^*$ was also discussed.
 Some kinematical properties of the signal were studied
(i.e.\ transverse mass and invariant mass distributions),
but in view of the sizeable backgrounds in these channels
a detailed simulation would be necessary to determine
whether the detection prospects were promising or not.

\subsection{Non-degenerate triplet scalars}
 The above decay modes are the dominant ones
for the case of degeneracy of the triplet scalars ($\lambda_4=0$)
or for the case of the mass hierarchy
where $H^0$ and $A^0$ are the lightest of the triplet scalars
(i.e.\ $m_{H^{\pm\pm}}^{} > m_{H^\pm}^{} > m_{H^0,A^0}^{}$
for $\lambda_4 < 0$).
 For the mass hierarchy
where $H^0$ and $A^0$ are the heaviest of the triplet scalars
($m_{H^{\pm\pm}}^{} < m_{H^\pm}^{} < m_{H^0,A^0}^{}$
for $\lambda_4 > 0$)
the decay channels $H^0\to H^\pm W^*$ and $A^0\to H^\pm W^*$
would be possible.
 The formulae for these decay widths can be easily obtained
from the analogous formula
for the decay width of $H^{\pm\pm}\to H^{\pm}W^*$.
 After summing over all fermion states
for  $W^*\to f^\prime \overline{f}$,
excluding the $t$ quark,
the decay rate is given by:
\begin{equation}
\Gamma(H^0\to H^{\pm} W^*\to  H^{\pm} f'\overline f)
\simeq
 \frac{ 9 G_F^2 m_W^4 m_{H^0}^{} }{ 8\pi^3 }
 \int_{0}^{1-\kappa_{H}^{}}\!\!\!dx_2
 \int_{1-x_2-\kappa_{H}^{}}^{1-\frac{\kappa_{H}^{}}{1-x_2}}
 \!\!\!dx_1 F_{HW}(x_1,x_2) ,
\label{HHWdecay}
\end{equation}
where $\kappa_{H}^{} \equiv m_{H^{\pm}}^{}/ m_{H^0}^{}$
and the analytical expression for $F_{ij}(x_1,x_2)$
can be found in Ref.~\cite{Djouadi:1995gv}
(see also Ref.~\cite{Moretti:1994ds}).
 Note that this decay mode does not depend on $v_\Delta$.
 In eq.~(\ref{HHWdecay})
we take $f'$ and $\overline f$ to be massless,
which is a good approximation
as long as the mass splitting
between $m_{H^0}^{}$ and $m_{H^\pm}^{}$ 
is above the mass of the charmed hadrons ($\simeq 2\,\GeV$).
 In our numerical analysis
we will be mostly concerned 
with such mass splittings, $m_{H^0}^{}-m_{H^{\pm}}^{}\gg 2\,\GeV$.
 It has already been shown that
the decays $H^{\pm\pm}\to H^{\pm}W^*$%
~\cite{Chakrabarti:1998qy,Chun:2003ej,
Akeroyd:2005gt,Perez:2008ha,Melfo:2011nx}
and $H^{\pm}\to H^{\pm\pm}W^*$~\cite{Akeroyd:2011zz}
can be the dominant decay channels
for the  charged scalars over a wide range of 
values of $v_\Delta$ and $\Delta M = (m_{H^{\pm}}^{}-m_{H^{\pm\pm}}^{}$),
even for $\Delta M \ll m_W$.
 Hence
we expect a similar result for the decays $H^0,A^0\to H^{\pm}W^*$,
which was mentioned (but not quantified) in Ref.~\cite{Akeroyd:2011zz}.

\begin{figure}[t]
\begin{center}
\includegraphics[origin=c, angle=-90, scale=0.32]{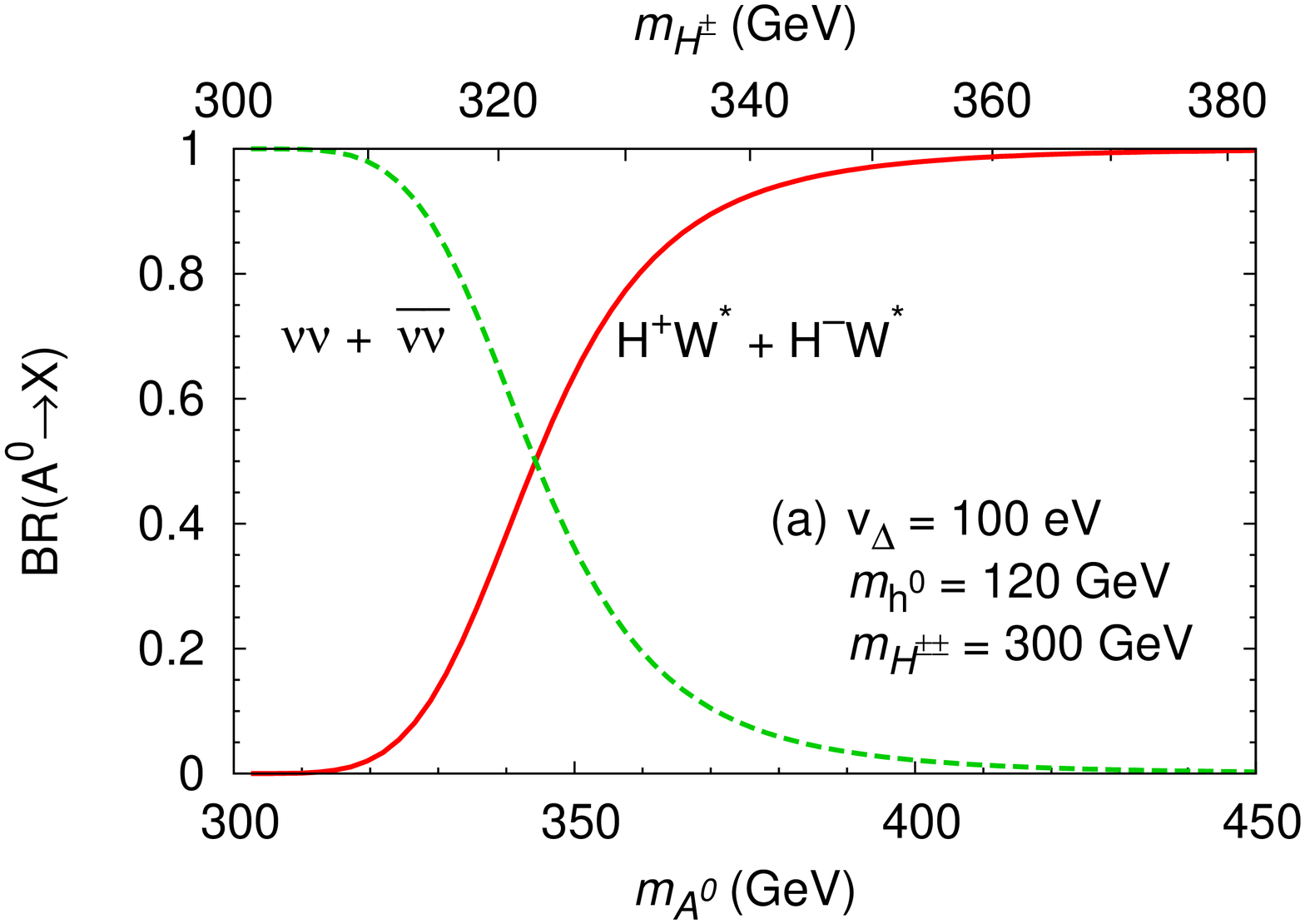}
\includegraphics[origin=c, angle=-90, scale=0.32]{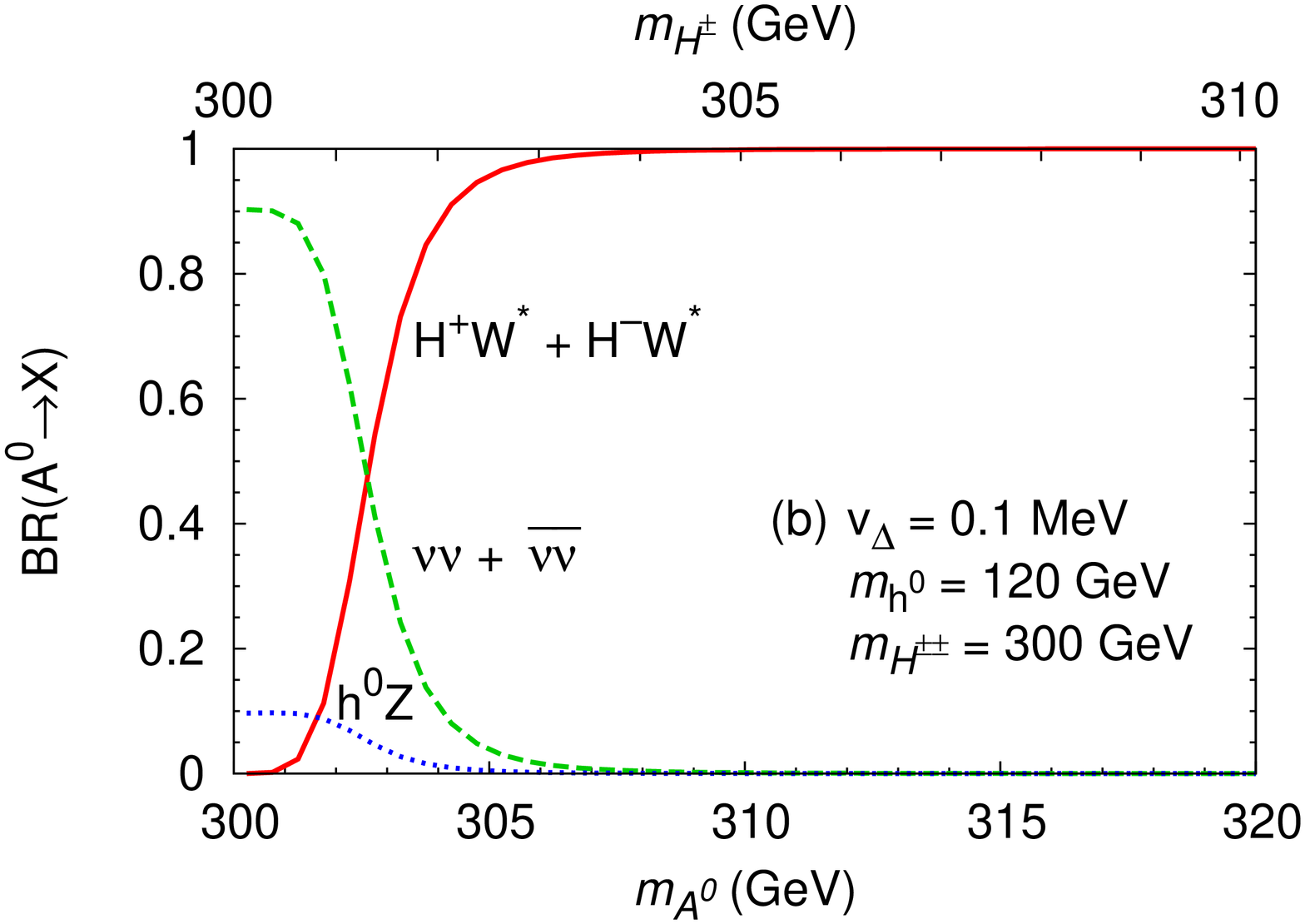}\\[-10mm]
\includegraphics[origin=c, angle=-90, scale=0.32]{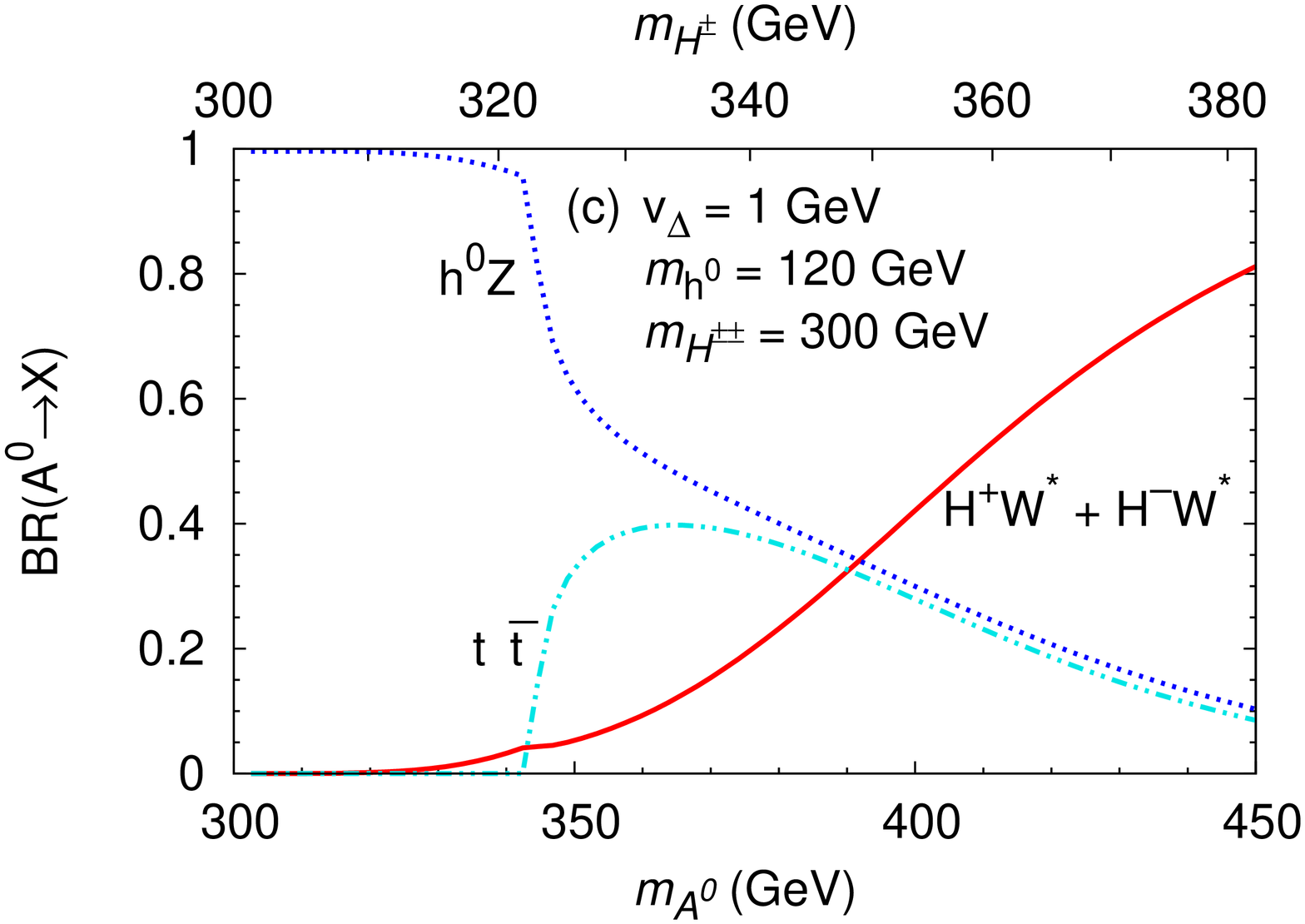}
\vspace*{-15mm}
\caption{
 BRs of $A^0$ as functions of $m_{A^0}^{}$
for (a) $v_\Delta=100\,\eV$,
(b) $v_\Delta=0.1\,\MeV$
and (c) $v_\Delta=1\,\GeV$.
 We fixed $m_{h^0}^{}=120\,\GeV$, $m_{H^{\pm\pm}}^{}=300\,\GeV$,
and $s_0^\prime = 2$.
The values of $m_{H^\pm}^{}$, obtained from the relation
$2 m_{H^\pm}^2 = m_{A^0,H^0}^2 + m_{H^{\pm\pm}}^2$,
are also shown.
The curves correspond to decays into
$H^+W^\ast + H^-W^\ast$ (red solid),
$\nu\nu + \overline{\nu}\,\overline{\nu}$ (green dashed),
$h^0Z$ (blue dotted),
and $t\overline{t}$ (cyan dot-dot-dashed).
}
\label{fig:br_A}
\end{center}
\end{figure}

 In Fig.~\ref{fig:br_A}
we show the BRs of $A^0$ as a function of $m_{A^0}$
for $v_\Delta=100\,\eV$, $v_\Delta=0.1\,\MeV$, and $v_\Delta=1\,\GeV$,
fixing $m_{H^{\pm\pm}}^{}=300\,\GeV$, $m_{h^0}^{}=120\,\GeV$,
and $s_0^\prime = 2$.
 We choose $m_{H^{\pm\pm}}^{}=300\,\GeV$
because this is the strongest experimental lower bound
on $m_{H^{\pm\pm}}^{}$,
which applies to the special case
of BR($H^{\pm\pm}\to \ell^\pm {\ell^\prime}^\pm)=100\%$
for $\ell=e,\mu$.
 Lower values of $m_{H^{\pm\pm}}^{}$ can be considered for the case of 
BR($H^{\pm\pm}\to \ell^\pm {\ell^\prime}^\pm)<100\%$
(assuming $\ell=e,\mu$)
or if the decays with $\ell=\tau$ dominate.
 A numerical analysis of the BRs of $A^0$ was
first presented in Ref.~\cite{Perez:2008ha} 
for the case of degeneracy of the triplet scalars (i.e.\ $\lambda_4=0$).
 In Ref.~\cite{Aoki:2011pz}
the case of $\lambda_4 < 0$ was considered,
in which $A^0$ (and $H^0$) 
is the lightest of the triplet scalars,
and so it has the same BRs as for the case
of degeneracy in Ref.~\cite{Perez:2008ha}.
 We present the first numerical study of the BRs of $A^0$
for the case of  $\lambda_4 > 0$,
for which the decay channel $A^0\to H^{\pm}W^*$ is open.
 The range of $m_{A^0}$ in the figures
corresponds to $0\leq \lambda_4 \lesssim 1$,
which easily satisfies the perturbative constraint $\lambda_4< 4\pi$.
 Very large mass splittings (e.g.\ $\gg 100\,\GeV$) are
constrained by measurements of electroweak precision observables.
 In Fig.~\ref{fig:br_A}(a) (for $v_\Delta=100\,\eV$)
one can see that $A^0\to H^+ W^\ast + H^- W^\ast$
competes with $H^0 \to \nu\nu + \overline{\nu}\overline{\nu}$,
and all other decay channels have a negligible BR\@.
 For ($m_{A^0}^{}-m_{H^{\pm\pm}}^{}) >50\,\GeV$,
$A^{0}\to H^+ W^\ast + H^- W^\ast$
becomes the dominant decay channel.
 The case of $v_\Delta\sim 0.1\,\MeV$ in Fig.~\ref{fig:br_A}(b)
gives the optimum results, 
where BR($A^0\to H^+ W^\ast + H^- W^\ast$) can reach $\sim 100\%$ 
even for small mass splittings ($m_{A^0}-m_{H^{\pm\pm}}) >5\,\GeV$.
 In Fig.~\ref{fig:br_A}(c) (for $v_\Delta=1\,\GeV$),
for which the competing decays
are $A^0\to h^0Z$ and $A^0\to t\overline t$,
the decay $A^0\to H^+W^\ast + H^- W^\ast$
becomes the dominant decay channel
for a mass splitting $(m_{A^0}-m_{H^{\pm\pm}}) >90\,\GeV$.

\begin{figure}[t]
\begin{center}
\includegraphics[origin=c, angle=-90, scale=0.32]{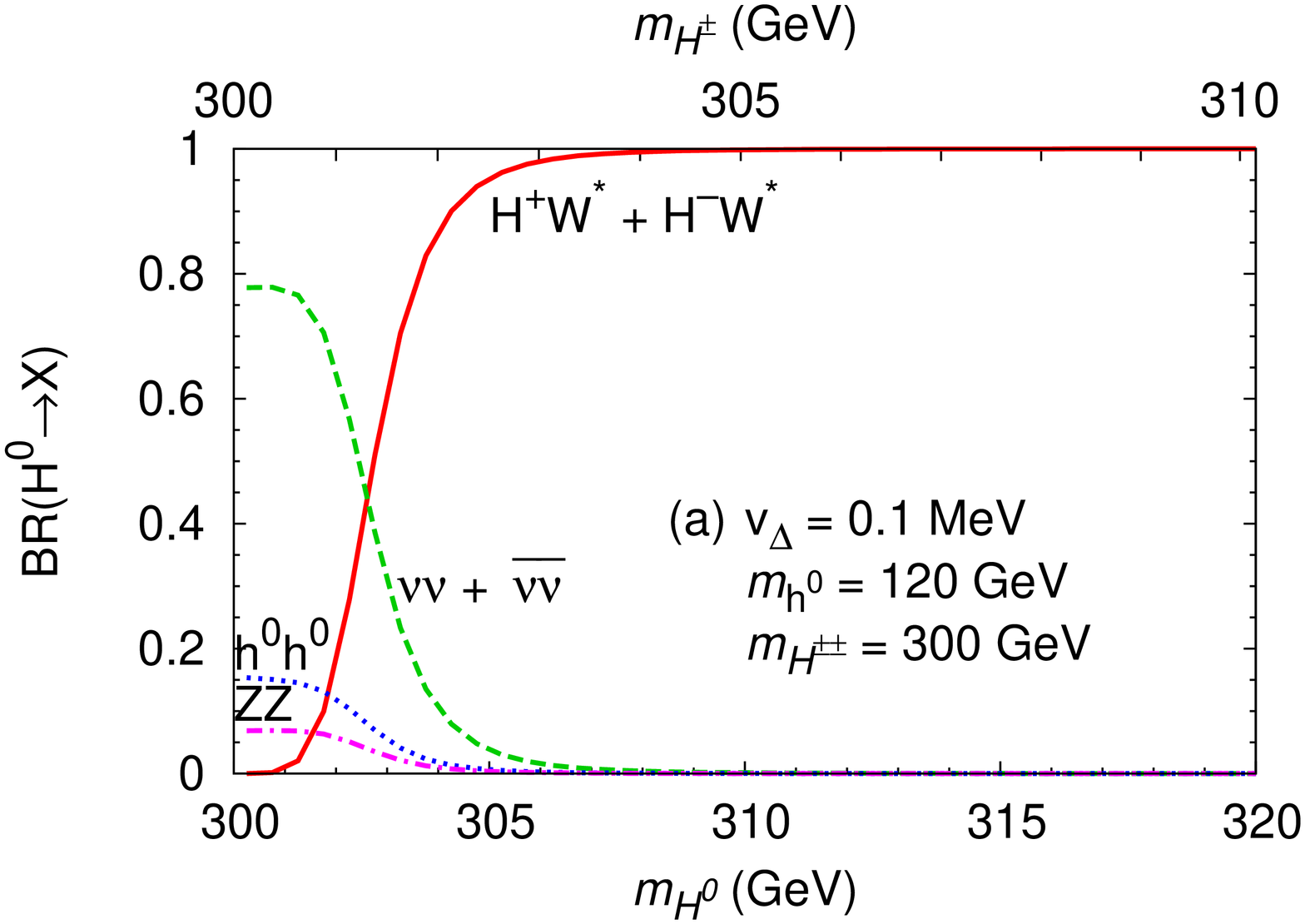}
\includegraphics[origin=c, angle=-90, scale=0.32]{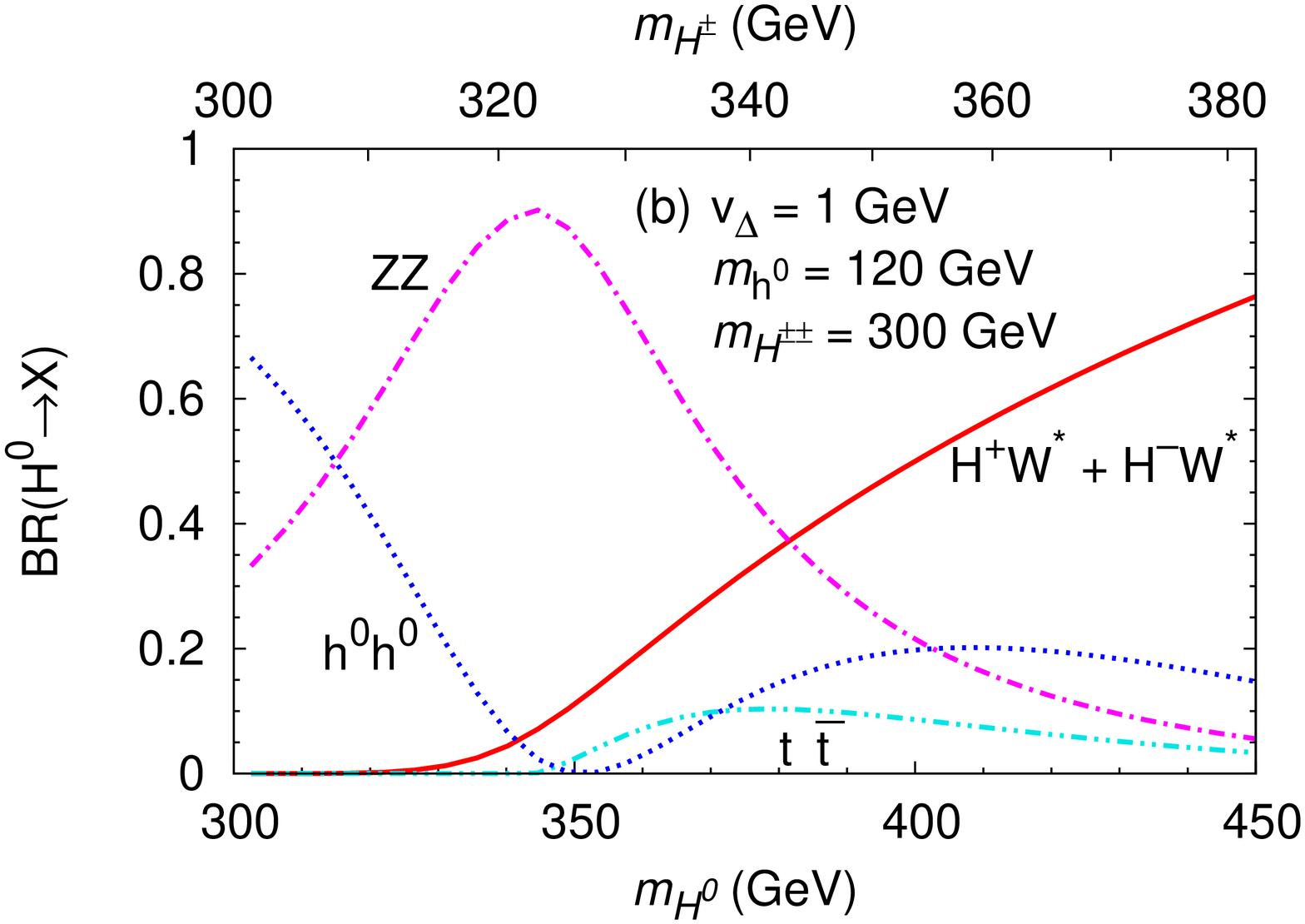}
\vspace*{-20mm}
\caption{
 BRs of $H^0$ as functions of $m_{H^0}^{}$
for (a) $v_\Delta=0.1\,\MeV$
and (b) $v_\Delta=1\,\GeV$.
 We fixed $m_{h^0}^{}=120\,\GeV$, $m_{H^{\pm\pm}}^{}=300\,\GeV$,
and $s_0^\prime = 2$. The values of $m_{H^\pm}^{}$, obtained from the relation
$2 m_{H^\pm}^2 = m_{A^0,H^0}^2 + m_{H^{\pm\pm}}^2$,
are also shown.
The curves correspond to decays into
$H^+W^\ast + H^-W^\ast$ (red solid),
$\nu\nu + \overline{\nu}\,\overline{\nu}$ (green dashed),
$h^0h^0$ (blue dotted),
$ZZ$ (magenta dot-dashed),
and $t\overline{t}$ (cyan dot-dot-dashed).
}
\label{fig:br_H}
\end{center}
\end{figure}

 In Fig.~\ref{fig:br_H}
we show the BRs of $H^0$,
again taking $m_{H^{\pm\pm}}^{}=300\,\GeV$, $m_{h^0}^{}=120\,\GeV$,
and $s_0^\prime = 2$.
 We do not show a figure for $v_\Delta = 100\,\eV$ since
it is identical to that of  Fig.~\ref{fig:br_A}(a).
 The optimal results for  BR($H^0\to H^+ W^\ast + H^- W^\ast$) 
are for $v_\Delta = 0.1\,\MeV$, and are shown in Fig.~\ref{fig:br_H}(b).
 For $v_\Delta = 1\,\GeV$
and approximate degeneracy $m_{H^0}^{}\sim m_{H^{\pm\pm}}^{}$
the dominant channel is $H^0\to h^0h^0$,
as pointed out in Ref.~\cite{Perez:2008ha}.
 As $m_{H^0}^{}$ increases up to $m_{H^0}^{}\sim 350\,\GeV$
(corresponding to an increasing value of $\lambda_4$),
the coupling $\lambda_{Hhh}$ decreases and vanishes
around $m_{H^0}^{}\sim 350\,\GeV$,
after which it increases in magnitude for $m_{H^0}^{}> 350\,\GeV$.
 This leads to a suppression of the partial width for $H^0\to h^0h^0$.
 Consequently,
$H^0\to ZZ$ becomes the dominant channel
in the mass range $320\,\GeV < m_{H^0}^{} < 380 \,\GeV$. 
 For $m_{H^0}^{}> 380\,\GeV$,
the decay $H^0\to H^+ W^\ast + H^- W^\ast$
is dominant.

 Importantly,
when BR($H^0\to H^{\pm}W^*$) is sizeable,
so too will be BR($H^\pm\to H^{\pm\pm}W^*$),
as shown in Ref.~\cite{Akeroyd:2011zz}.
 Therefore
the decay mode $H^0\to H^{\pm}W^*$
would lead to production of $H^{\pm\pm}$
accompanied by $W^*W^*$,
which would provide a promising experimental signature.
 We will quantify the magnitude of
the cross section of this signature in the next section.

\section{Numerical Analysis} 
 In this section
we quantify the magnitude
of the number of pair-produced $H^{++}H^{--}$
arising from the production of $H^0$ and $A^0$
with decay $H^0,A^0\to  H^{\pm}W^*$.
 The important parameters for our analyses are
$v_\Delta^{}$, $m_{A^0}^{} (\simeq m_{H^0}^{})$, and $m_{H^{\pm\pm}}^{}$ 
(with which $m_{H^\pm}^2 \simeq (m_{A^0}^2 + m_{H^{\pm\pm}}^2)/2$).
 We present results for leading-order cross sections
using CTEQL1 parton distribution functions~\cite{Pumplin:2002vw},
with the scale taken to be
the partonic centre-of-mass energy ($Q=\hat s$).

 In Ref.~\cite{Akeroyd:2011zz}
the cross sections $X_3$ and $X_3^\prime$ 
were defined as follows:
\begin{eqnarray}
X_3
&\equiv&
 \left\{
  \sigma(\ppHpH) + \sigma(\ppHmH)
 \right\}
\nonumber\\
&&\hspace*{50mm}
\times
 \BR_+\,
 [\BR(H^\pm \to H^{\pm\pm} W^*)]^2 ,
\label{eq:HpH0}
\\
%
%
X_3^\prime
&\equiv&
 \left\{
  \sigma(\ppHpH) + \sigma(\ppHmH)
 \right\}
\nonumber\\
&&\hspace*{50mm}
\times
 \BR_-\,
 [\BR(H^\pm \to H^{\pm\pm} W^*)]^2 ,
\label{eq:HpH0-2}
\\
%
%
\BR_\pm
&\equiv&
 \BR(H^0\to H^\pm W^*)
 +
 \BR(A^0\to H^\pm W^*)
\nonumber\\
&&\hspace*{20mm}
{}\pm
 \frac{ 4 \BR(H^0\to H^\pm W^*) \BR(A^0\to H^\pm W^*) }
      { \BR(H^0\to H^\pm W^*) + \BR(A^0\to H^\pm W^*) } ,
\end{eqnarray}
where we used
$\sigma(\ppHpmA) \simeq \sigma(\ppHpmH)$
because $m_{A^0} \simeq m_{H^0}$.

 The signature arising from $X_3$ is:
\begin{equation}
H^{++}H^{--} \;\;{\rm with}\;\; W^{\pm *} W^{\pm *} W^{\mp *}\;\;
\end{equation}
 In this process $L\#$ is conserved.
 In contrast,
the signature arising from $X_3^\prime$ is:
\begin{equation}
H^{++}H^{++} \;\;({\rm with} \;3(W^-)^\ast)
\;\;{\rm and} \;\; H^{--}H^{--} \;\; ({\rm with}\; 3(W^+)^\ast) .
\end{equation}
 In this process $L\#$ is violated,
and so its magnitude is much smaller than that of $X_3$.
 In the limit of  $L\#$ conservation (i.e.\ $v_\Delta\to 0$)
one has $\Gamma_{\text{tot}}(H^0) = \Gamma_{\text{tot}}(A^0)$ and
BR$(A^0\to H^\pm W^*)$=BR$(H^0\to H^\pm W^*)$.
 Consequently,
$X_3'\to 0$ as $v_\Delta\to 0$.
 The quantity $X_3+X_3'$ is given by the simple expression:
\begin{eqnarray}
X_3+X_3'
&=&
 2
 \left\{
  \sigma(\ppHpH) + \sigma(\ppHmH)
 \right\}
\nonumber\\
&&\hspace*{0mm}
 \times
 \Bigl[
  \BR(H^0\to H^\pm W^*)
  +
  \BR(A^0\to H^\pm W^*)
 \Bigr]
 [\BR(H^\pm \to H^{\pm\pm} W^*)]^2 .
\end{eqnarray}  
 The $W^*W^*W^*$ from the cascade decays lead to additional leptons,
quarks and missing energy.
 For $\ell=e,\mu$,
$49.8\%$ of the final states from $W^*W^*W^*$ have $5\ell$ or more,
with $9.6\%$ having $6\ell$ and $0.8\%$ having $7\ell$.
 Of particular interest are the final states
with $5\ell$ and $6\ell$,
due to their small backgrounds and their similarity
with the ongoing searches for $4\ell$.
 We define their cross sections (with $\ell=e,\mu$) as follows:
\begin{equation}
X^{5\ell}_3
=
 0.498(X_3+X_3^\prime) ,\;\;
%
X^{6\ell}_3
=
 0.096(X_3+X_3^\prime)\,.
\end{equation}
 The decay mode $A^0 \to h^0 Z\to Z^*ZZ$
can give rise to $5\ell$ and $6\ell$ signatures,
but its magnitude is very suppressed
by the third power of BR($Z\to \ell\ell$),
and small BR($h^0\to Z^*Z$) for $m_{h^0}=120\,\GeV$ of interest to us.

\begin{figure}[t]
\begin{center}
\includegraphics[origin=c, angle=-90, scale=0.32]{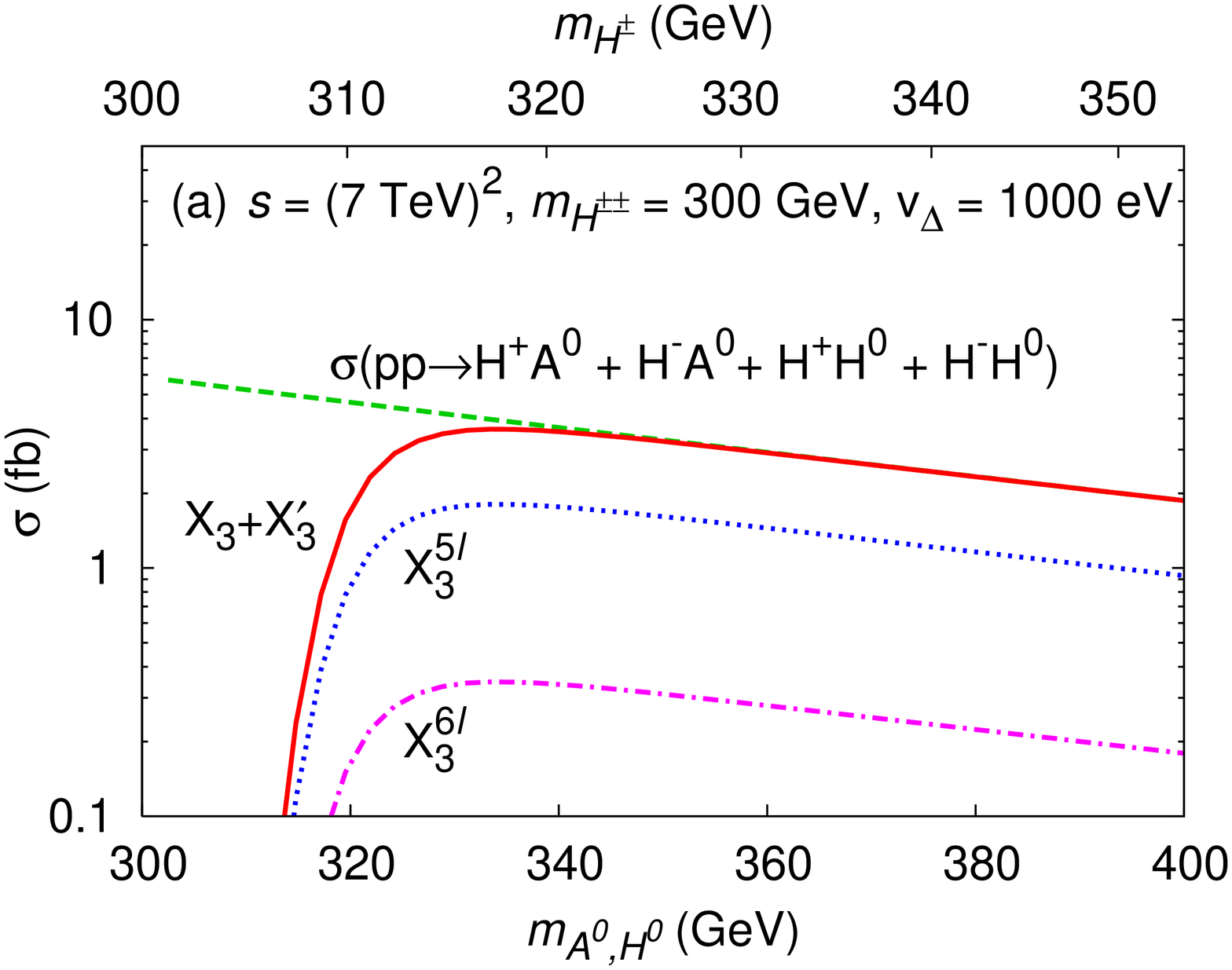}
\includegraphics[origin=c, angle=-90, scale=0.32]{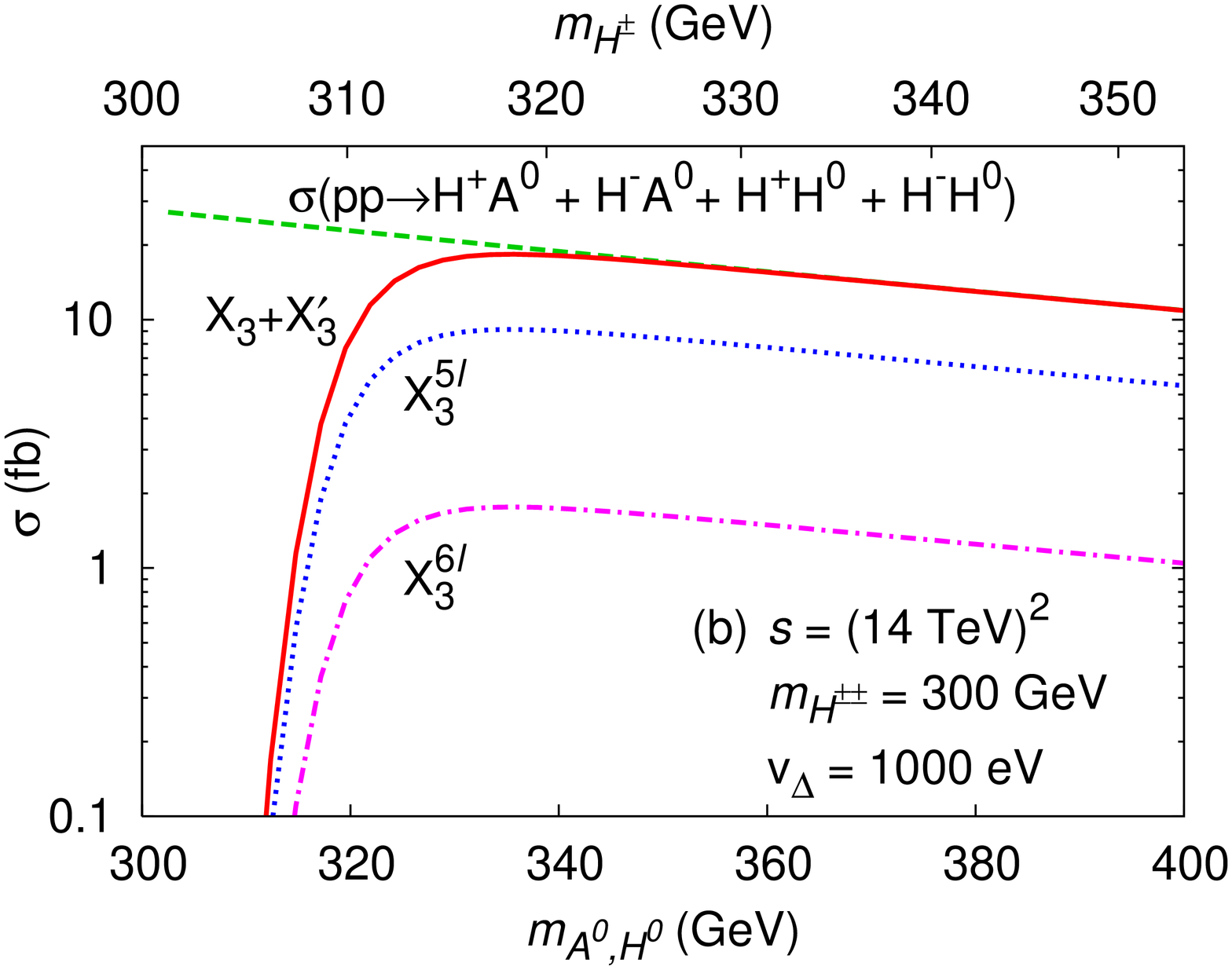}
\vspace*{-15mm}
\caption{
 The cross sections
for (a) $\sqrt s=7\,\TeV$
and (b) $\sqrt s=14\,\TeV$
at the LHC\@.
 We fixed $m_{h^0}^{}=120\,\GeV$, $m_{H^{\pm\pm}}^{}=300\,\GeV$,
$v_\Delta = 1000\,\eV$, and $s_0^\prime = 2$.
The values of $m_{H^\pm}^{}$, obtained from the relation
$2 m_{H^\pm}^2 = m_{A^0,H^0}^2 + m_{H^{\pm\pm}}^2$,
are also shown.
The green dashed curve shows
$\sigma(pp\to H^+ A^0 + H^- A^0 + H^+ H^0 + H^- H^0)$.
Other curves correspond to
cascade decays ($X_3$ + $X_3^\prime$, red solid),
$5\ell$ signature ($X_3^{5\ell}$, blue dotted),
and $6\ell$ signature ($X_3^{6\ell}$, magenta dot-dashed). 
}
\label{fig:X3}
\end{center}
\end{figure}

 The magnitude of $X_3+X_3'$ was not studied
in our earlier work in Ref.~\cite{Akeroyd:2011zz}.
 In Fig.~\ref{fig:X3}
we plot the quantities
$\sigma(pp\to H^+A^0 + H^-A^0 + H^+H^0 + H^-H^0)$,
$X_3+X_3^\prime$, $X^{5\ell}_3$
and $X^{6\ell}_3$ as functions of $m_{A^0}^{}$,
fixing $m_{H^{\pm\pm}}=300\,\GeV$ and $v_\Delta=1000\,\eV$,
and for $\sqrt{s}=7\,\TeV$ and $\sqrt{s}=14\,\TeV$.
 For this choice of $v_\Delta=1000\,\eV$,
a mass splitting $m_{A^0}^{}-m_{H^{\pm\pm}}^{}$ of around $40\,\GeV$
gives BR$(A^0\to H^\pm W^*) \simeq $BR$(H^0\to H^\pm W^*) \simeq 50\%$,
and thus
$X_3+X_3'\simeq 2 \sigma(pp\to H^+H^0 + H^-H^0)$.
 The magnitude of $X^{5\ell}_3$ can reach $2\,\fb$ and $9\,\fb$
for $\sqrt{s}=7\,\TeV$ and  $\sqrt{s}=14\,\TeV$
respectively, while $X^{6\ell}_3$ can reach $0.3\,\fb$ and $2\,\fb$.
 It is likely that up to $20\,\fb^{-1}$ 
will be accumulated during the run of the LHC at $\sqrt{s}=7\,\TeV$,
and therefore $X^{5\ell}_3 \simeq 2\,\fb$ would give up to $40$ $5\ell$ events
and $X^{6\ell}_3$ would give up to 6 $6\ell$ events
before selection cuts.
 We note that these event numbers
need to be multiplied by the the square of
$\sum\BR(H^{\pm\pm} \to e^\pm e^\pm, e^\pm \mu^\pm, \mu^\pm\mu^\pm)$,
which we take to be $100\%$ for illustration.
 In the HTM
the maximum value of this sum is $\simeq 70\%$%
~\cite{Garayoa:2007fw,Akeroyd:2007zv,Kadastik:2007yd,
Perez:2008ha,delAguila:2008cj,Akeroyd:2009hb}, 
and so the above cross sections would be
multiplied by $\simeq 0.5$ in the optimal case. 
 Note that our numerical analysis is for $m_{H^{\pm\pm}}^{}= 300\,\GeV$.
 Values of $m_{H^{\pm\pm}}^{}< 300\,\GeV$
(which would allow lighter $m_{H^\pm}^{}$, $m_{A^0}^{}$ and $m_{H^0}^{}$)
can be considered,
provided that $\BR(H^{\pm\pm} \to \ell^\pm {\ell^\prime}^\pm)<100\%$
in order to respect the current limits on $m_{H^{\pm\pm}}^{}$.
 However,
due to this latter suppression factor,
the magnitudes of $X^{5\ell}_3$ and $X^{6\ell}_3$
for $m_{H^{\pm\pm}}^{}< 300\,\GeV$
are no larger than the maximum cross sections
for the case of $m_{H^{\pm\pm}}= 300\,\GeV$.
 The search by the CMS collaboration~\cite{CMS-search}
has presented mass limits
of $m_{H^{\pm\pm}}^{} > 269\,\GeV, 297\,\GeV, 291\,\GeV$ and $289\,\GeV$
for four benchmark points of BR($H^{\pm\pm}\to \ell^\pm{\ell^\prime}^\pm$)
in the HTM 
(for the scenario of interest of $v_\Delta < 0.1\,\MeV$).
 In light of this,
the choice of $m_{H^{\pm\pm}}^{}=300\,\GeV$ is an appropriate one
for quantifying the largest attainable values
of $X^{5\ell}_3$ and $X^{6\ell}_3$,
while satisfying the experimental bounds on $m_{H^{\pm\pm}}^{}$.

 The backgrounds at the LHC (with $\sqrt{s}=14\,\TeV$ and $30\,\fb^{-1}$)
for the $5\ell$ and $6\ell$ signatures have been evaluated
in Ref.~\cite{delAguila:2008cj}.
 The background for the $6\ell$ signature was shown to be negligible 
after requiring two of the leptons to have a transverse momentum~($p_T$)
larger than $30\,\GeV$,
and  the remaining leptons to have $p_T>15\,\GeV$.
 Employing the same cuts for the $5\ell$ background,
around 1 event was shown to remain
for an integrated luminosity of $30\,\fb^{-1}$.
 Therefore
the $5\ell$ and $6\ell$ signatures have the virtue of having no background.
 We expect this conclusion to be also true for the LHC with $\sqrt{s}=7\,\TeV$
and a final integrated luminosity of $20\,\fb^{-1}$.

\begin{figure}[t]
\begin{center}
\includegraphics[origin=c, angle=-90, scale=0.32]{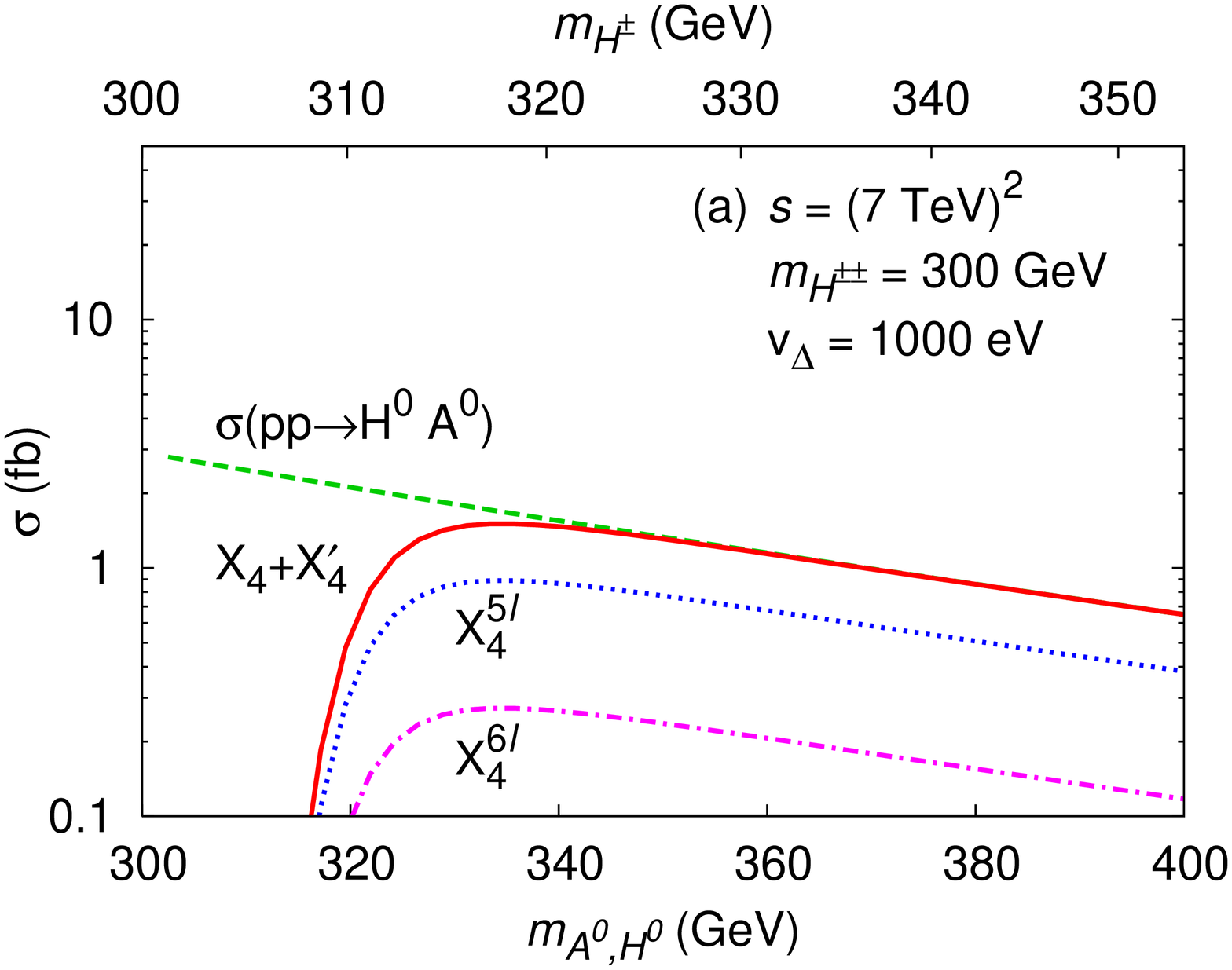}
\includegraphics[origin=c, angle=-90, scale=0.32]{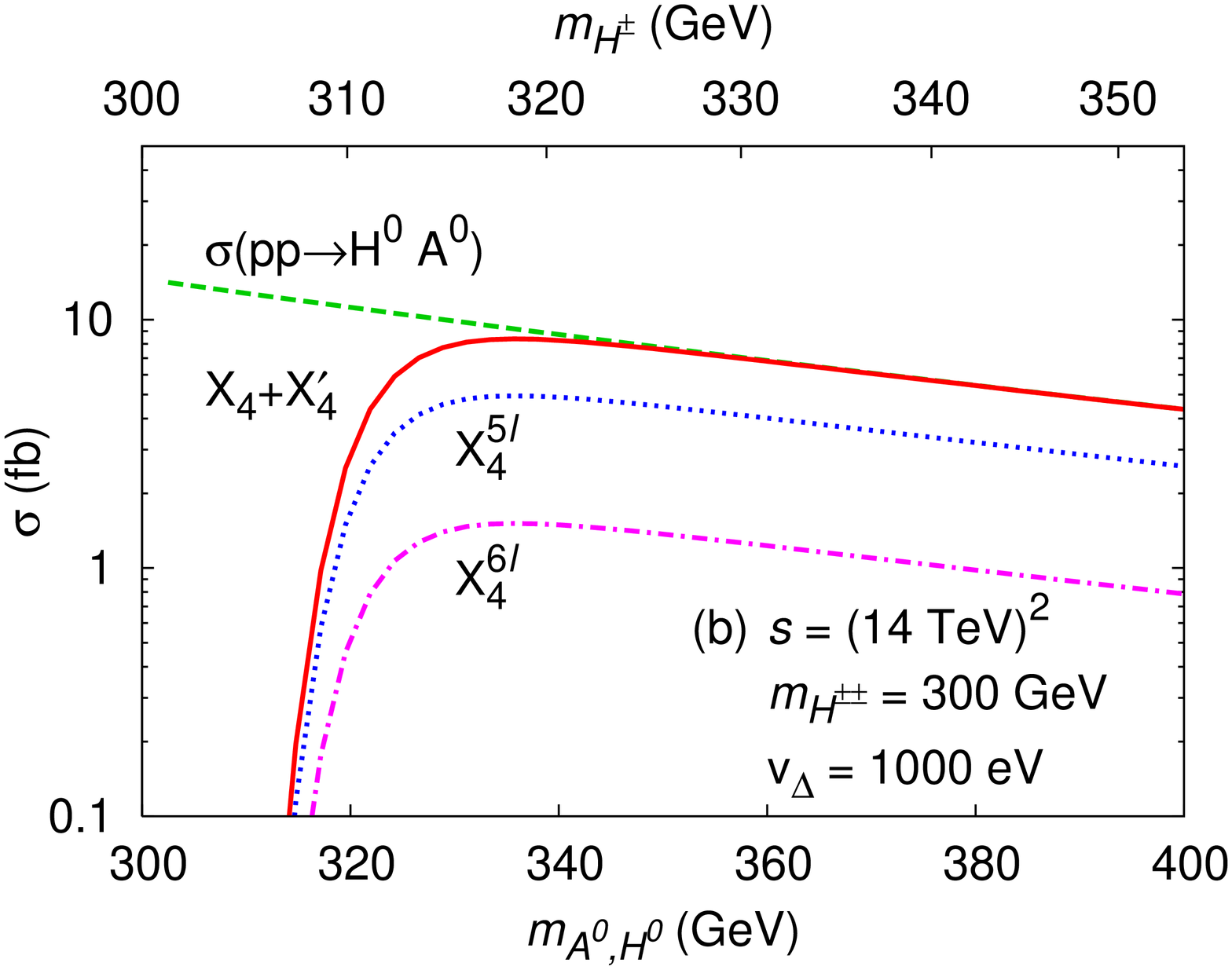}
\vspace*{-15mm}
\caption{
 The cross sections
for (a) $\sqrt s=7\,\TeV$
and (b) $\sqrt s=14\,\TeV$
at the LHC\@.
We fixed $m_{h^0}^{}=120\,\GeV$, $m_{H^{\pm\pm}}^{}=300\,\GeV$,
$v_\Delta = 1000\,\eV$, and $s_0^\prime = 2$.
The values of $m_{H^\pm}^{}$, obtained from the relation
$2 m_{H^\pm}^2 = m_{A^0,H^0}^2 + m_{H^{\pm\pm}}^2$,
are also shown.
The green dashed curve shows $\sigma(pp\to H^0 A^0)$.
 Other curves correspond to
cascade decays ($X_4$ + $X_4^\prime$, red solid),
$5\ell$ signature ($X_4^{5\ell}$, blue dotted),
and $6\ell$ signature ($X_4^{6\ell}$, magenta dot-dashed).
}
\label{fig:X4}
\end{center}
\end{figure}

 There is another production mechanism which scales as $\BR^4$,
and involves the production of neutral scalars only%
\footnote{
 Expressions of $X_4$ and $X_4^\prime$ in Ref.~\cite{Akeroyd:2011zz}
were not correct
although they were not used for analyses in the article.
}
:
\begin{eqnarray}
X_4
&\equiv&
 \frac{1}{\,4\,}
 \sigma(\ppHA)\,
 [\BR_+ - \BR_-]
 \BR_+\,
 [\BR(H^\pm\to H^{\pm\pm} W^*)]^2 ,
\\
X_4^\prime
&\equiv&
 \frac{1}{\,4\,}
 \sigma(\ppHA)\,
 [\BR_+ - \BR_-]
 \BR_-\,
 [\BR(H^\pm\to H^{\pm\pm} W^*)]^2 ,
\end{eqnarray}
\begin{eqnarray}
X_4 + X_4^\prime
&=&
 4 \sigma(\ppHA)
\nonumber\\
&&\hspace*{10mm}
 \times
 \BR(H^0\to H^\pm W^*)\,
 \BR(A^0\to H^\pm W^*)\,
 [\BR(H^\pm\to H^{\pm\pm} W^*)]^2 .
\end{eqnarray}
 The signature arising from $X_4$ is:
\begin{equation}
H^{++}H^{--} \;\;{\rm with}\;\; W^{+ *} W^{+ *} W^{- *}W^{- *}\,.
\end{equation}
 The signature arising from $X_4^\prime$ is:
\begin{equation}
H^{++}H^{++} \;\;{\rm with}\;\; 4W^{-*}  \;\;{\rm and} \;\;H^{--}H^{--} \;\;{\rm with}\;\; 4W^{+*}\,.
\end{equation}
 In contrast to $X_4$,
the mechanism $X_4^\prime$ gives a pair of {\it same-sign} $H^{\pm\pm}$
(being proportional to $\BR_-$, like $X_3^\prime$)
and its magnitude is negligible for small $v_\Delta$.
 The $W^*W^*W^*W^*$ from the cascade decays lead to additional leptons,
quarks and missing energy.
 For $\ell=e,\mu$,
$59.04\%$ of the final states from $W^*W^*W^*W^*$ have $5\ell$ or more,
with $18.08\%$ having $6\ell$ and 2.72\% having $7\ell$ or more.
 We define the cross sections for production of $5\ell$ and $6\ell$
(with $\ell=e,\mu$) as follows:
\begin{equation}
X^{5\ell}_4
=
 0.5904(X_4+X_4^\prime)\,,\;\;
%
X^{6\ell}_4
=
 0.1808(X_4+X_4^\prime)\,.
\end{equation}
 Figure~\ref{fig:X4} is similar to Fig.~\ref{fig:X3}
but is for $X_4+X_4^\prime$. 
 The magnitude of $X^{5\ell}_4$ can reach $0.9\,\fb$ and $5\,\fb$
for $\sqrt{s}=7\,\TeV$ and  $\sqrt{s}=14\,\TeV$
respectively, while $X^{6\ell}_4$ can reach $0.3\,\fb$ and $1.5\,\fb$.

\begin{figure}[t]
\begin{center}
\includegraphics[origin=c, angle=-90, scale=0.32]{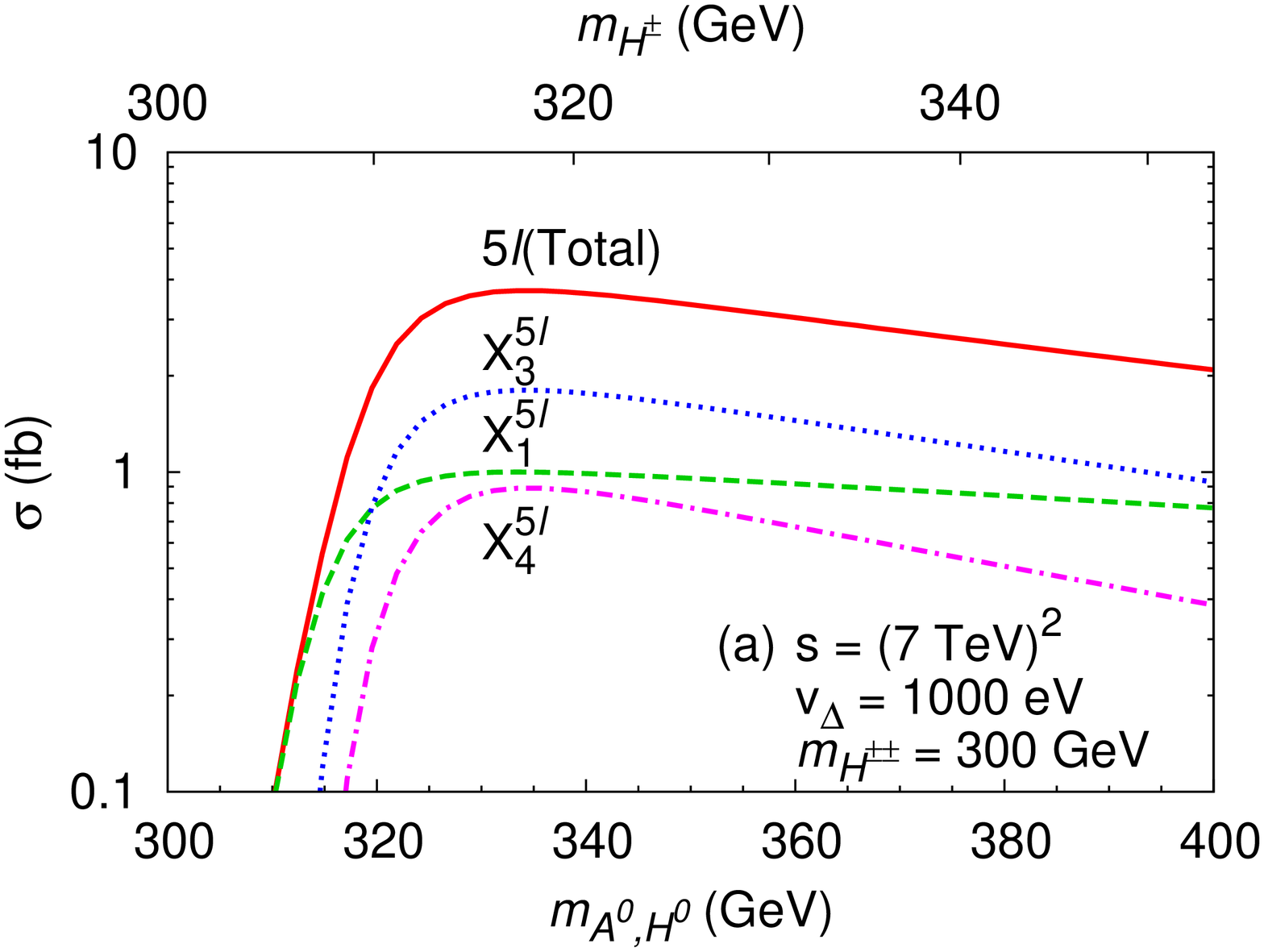}
\includegraphics[origin=c, angle=-90, scale=0.32]{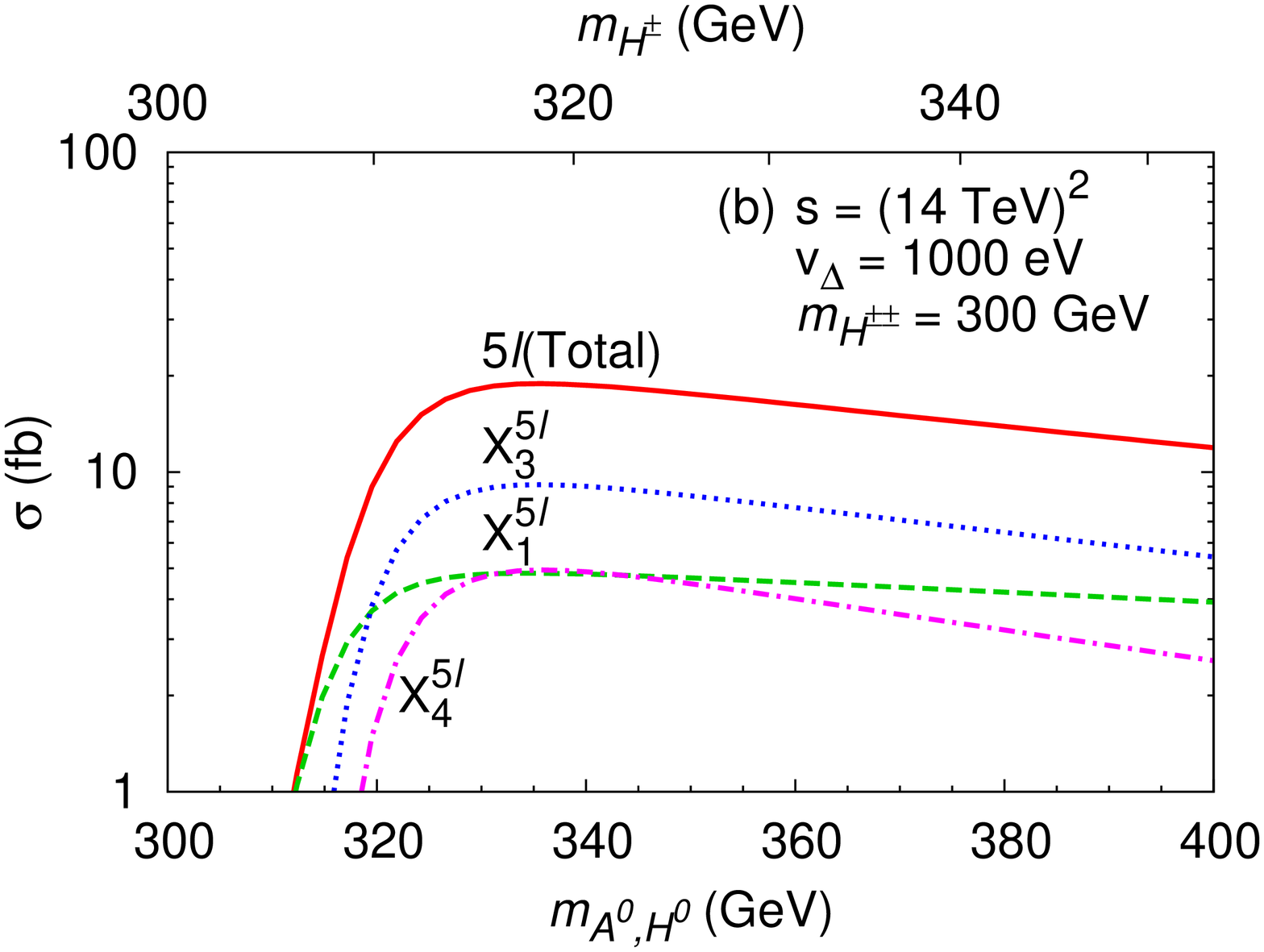}
\vspace*{-20mm}
\caption{
 The five-lepton cross sections
for (a) $\sqrt{s}=7\,\TeV$ and (b) $\sqrt{s}=14\,\TeV$
at the LHC\@.
 We fixed $m_{h^0}^{}=120\,\GeV$, $m_{H^{\pm\pm}}^{}=300\,\GeV$,
$v_\Delta = 1000\,\eV$, and $s_0^\prime = 2$.
The values of $m_{H^\pm}^{}$, obtained from the relation
$2 m_{H^\pm}^2 = m_{A^0,H^0}^2 + m_{H^{\pm\pm}}^2$,
are also shown.
 Contributions from
$H^{\pm\pm} H^\mp$ ($X_1^{5\ell}$, green dashed),
$H^\pm H^0 (A^0)$ ($X_3^{5\ell}$, blue dotted),
and $H^0 A^0$ ($X_4^{5\ell}$, magenta dot-dashed)
are shown individually, and the red curve is for their sum. 
}
\label{fig:5l}
\end{center}
\end{figure}

\begin{figure}[t]
\begin{center}
\includegraphics[origin=c, angle=-90, scale=0.32]{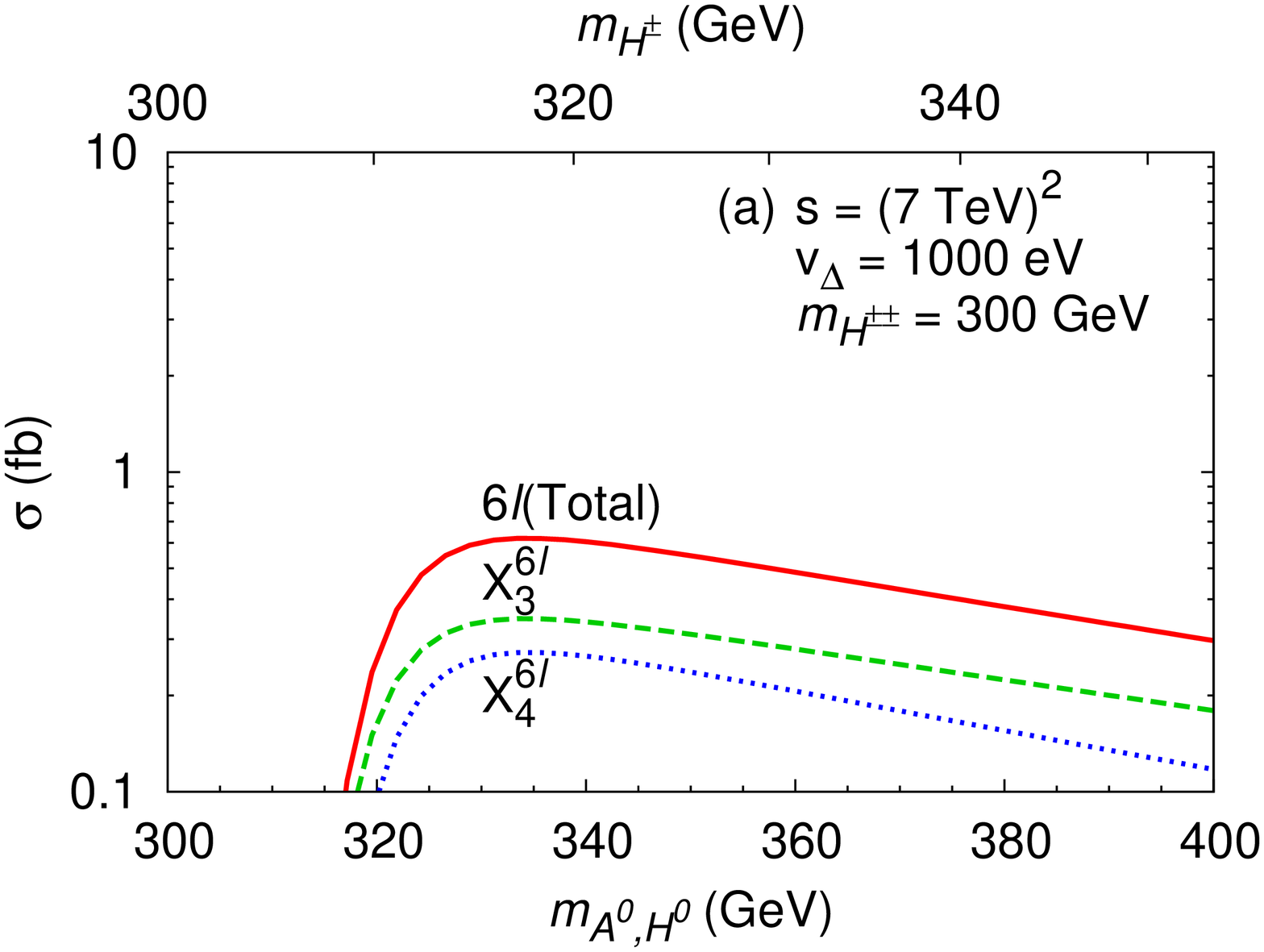}
\includegraphics[origin=c, angle=-90, scale=0.32]{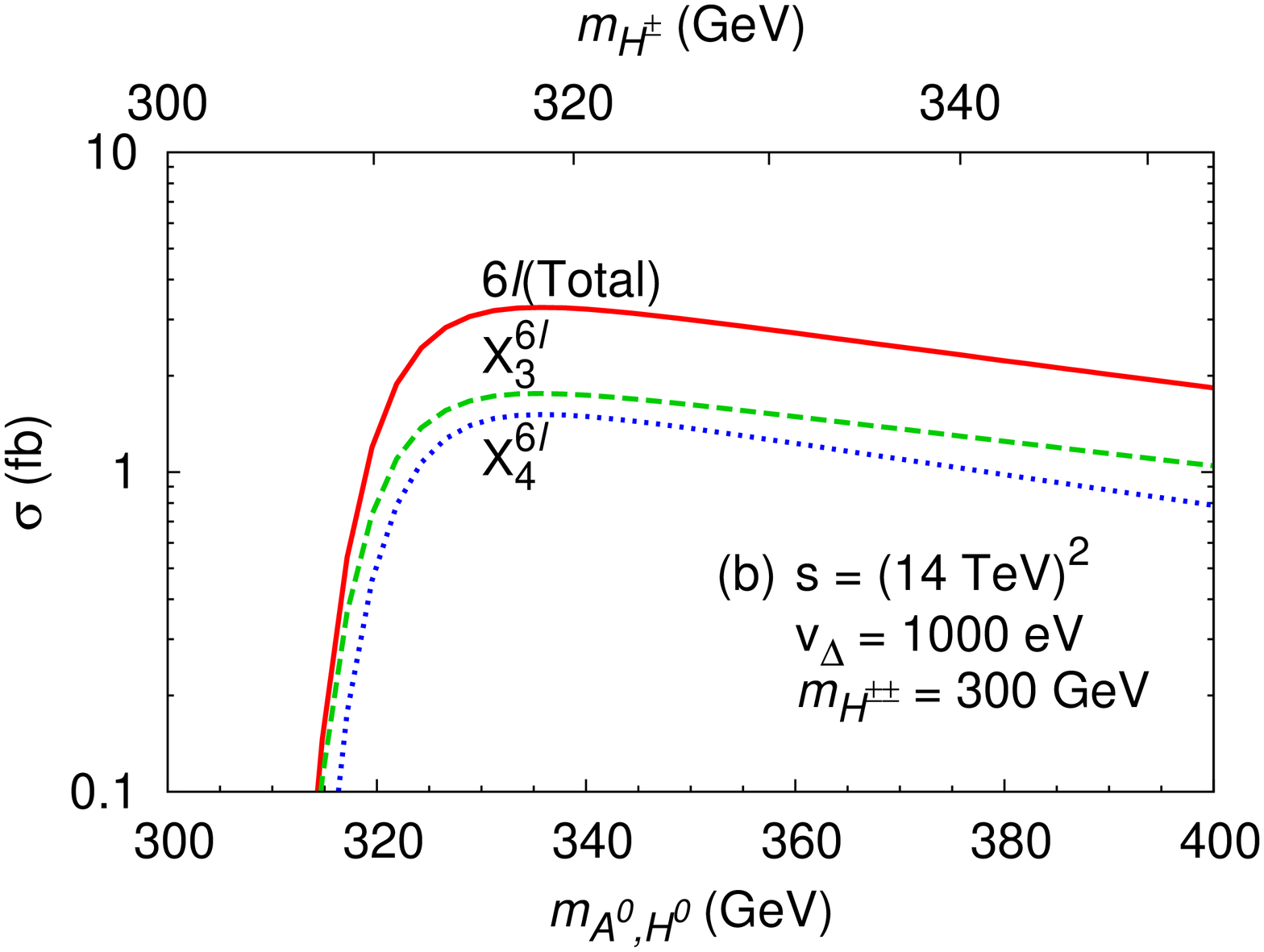}
\vspace*{-20mm}
\caption{
 The six-lepton cross sections
for (a) $\sqrt{s}=7\,\TeV$ and (b) $\sqrt{s}=14\,\TeV$
at the LHC\@.
 We fixed $m_{h^0}^{}=120\,\GeV$, $m_{H^{\pm\pm}}^{}=300\,\GeV$,
$v_\Delta = 1000\,\eV$, and $s_0^\prime = 2$.
The values of $m_{H^\pm}^{}$, obtained from the relation
$2 m_{H^\pm}^2 = m_{A^0,H^0}^2 + m_{H^{\pm\pm}}^2$,
are also shown.
The contributions from
$H^\pm H^0 (A^0)$ ($X_3^{6\ell}$, green dashed)
and $H^0 A^0$ ($X_4^{6\ell}$, blue dotted)
are shown individually, and the red curve is for their sum.
}
\label{fig:6l}
\end{center}
\end{figure}

 We now compare the magnitudes of the above distinct production mechanisms
for the $5\ell$ and $6\ell$ signatures,
and study the total cross section in these channels.
 There are three different mechanisms
which give rise to the $5\ell$ signature:
 $X^{5\ell}_1$ (which was studied in Ref.~\cite{Akeroyd:2011zz}),
 $X^{5\ell}_3$ and $X^{5\ell}_4$ (studied above).
 The individual contributions and their sum are plotted in Fig.~\ref{fig:5l}.
 One can see that $X^{5\ell}_3$ gives the largest cross section,
largely due to the combinatorial factor of 0.498
compared to 0.2 for $X_1$.
 Note that the $X^{5\ell}_1$ decreases more slowly
than $X^{5\ell}_3$ and $X^{5\ell}_4$ as $m_{A^0,H^0}^{}$ increases.
 This is because
the mass splitting $m_{H^\pm}^{}-m_{H^{\pm\pm}}^{}$ is relevant
for $X^{5\ell}_1$,
while $m_{A^0,H^0}^{}-m_{H^{\pm}}^{}$ is relevant
for $X^{5\ell}_3$ and $X^{5\ell}_4$.
 One can see that
the total cross section for $5\ell$ can reach
$4\,\fb$ for $\sqrt{s}=7\,\TeV$ and $20\,\fb$ for $\sqrt{s}=14\,\TeV$.
 The inclusion of the QCD $K$ factor of about 1.3~\cite{Dawson:1998py}
would increase these cross sections to $5\,\fb$ and $25\,\fb$ respectively.
 In Fig.~\ref{fig:6l}
we plot both $X^{6\ell}_3$ and $X^{6\ell}_4$.
 Note that $X_1$ cannot give a $6\ell$ signature.
 The total cross section for $6\ell$ can reach
$0.6\,\fb$ for $\sqrt{s}=7\,\TeV$ and $3\,\fb$ for $\sqrt{s}=14\,\TeV$.
 With the $K$ factor,
these cross sections become $0.8\,\fb$ and $4\,\fb$.

 To determine whether cross sections of these magnitudes
would be observable at the LHC
we now discuss the typical signal efficiencies
for the ongoing searches for $H^{\pm\pm}\to \ell^\pm{\ell^\prime}^\pm$
at the LHC\@.
 In the CMS search~\cite{CMS-search} for four leptons ($4\ell$)
one has approximate signal efficiencies of
$58\%$ for $eeee$, $76\%$ for $ee\mu\mu$
and $\sim 100\%$ for the $\mu\mu\mu\mu$ channels,
with a dependence on $m_{H^{\pm\pm}}^{}$.
 For the $5\ell$ and $6\ell$ channels
the requirement of one and two additional leptons respectively
would cause a drop in efficiency,
especially because the extra lepton has come from $W^*$
and so it would be less energetic than leptons
which originate from $H^{\pm\pm}$.
 However,
given the relatively high efficiencies
(especially for signature $\mu\mu\mu\mu$) in the $4\ell$ channel
we conclude that the $5\ell$ signature with a cross section as much as $5\,\fb$
has a good chance of being observed at the LHC with $\sqrt{s}=7\,\TeV$.
 We defer a detailed study of the decrease in efficiency
for the $5\ell$ and $6\ell$ channels
(relative to that for the $4\ell$ channel) to a future work.
 We note that the majority of the $5\ell$ signal
comes from the production processes
which involve  neutral Higgs bosons ($X^{5\ell}_3$ and $X^{6\ell}_4$),
while $6\ell$ can only originate from the production mechanisms
which involve neutral Higgs bosons.

 Finally,
we mention that such $5\ell$ and $6\ell$ signatures
can arise in other models
e.g.\ the Type~III seesaw mechanism~\cite{Foot:1988aq},
where $6\ell$ originates from pair-production of a heavy lepton $E^+E^-$
followed by decay $E^\pm \to \ell^\pm Z$ and $Z\to \ell^+\ell^-$,
and $5\ell$ comes from production of $E^\pm$
in association with a heavy neutrino $N$.
 The magnitude of the signal in the Type~III seesaw
was studied in Ref.~\cite{delAguila:2008cj},
and it was shown that for $m_{E^\pm}^{}=300\,\GeV$
(i.e.\ the same mass as $m_{H^{\pm\pm}}^{}$ in our numerical analysis)
less than 1 signal event would survive the above $p_T$ cuts
for the $6\ell$ channel with $30\,\fb^{-1}$.
 Prospects are better for the $5\ell$ channel,
and a significant signal (10 events) could observed
in the Type III seesaw with $30\,\fb^{-1}$ with $\sqrt{s}=14\,\TeV$.
 We note that
these event numbers are likely to be considerably smaller
than the analogous numbers for the HTM with $m_{H^{\pm\pm}}^{}=300\,\GeV$,
one reason being that
the extra leptons originate from $W^*\to \ell\nu$ (BR=$20\%$) in the HTM,
while in the Type~III seesaw they originate from $Z\to \ell^+\ell^-$ (BR=$6\%$). To illustrate this,
we recall that a total cross section of $20\,\fb$ is possible
for the $5\ell$ channel in the HTM with $\sqrt{s}=14\,\TeV$,
which would give 600 $5\ell$ events
{\it before} selection cuts for a luminosity of $30\,\fb^{-1}$. 
 Therefore
only a relatively small signal efficiency ($>1/60$)
would be needed to have $> 10$ $5\ell$ events.
 Moreover,
the $6\ell$ signature in the HTM with a cross section of $3\,\fb$
would give $>10$ events with $30\,\fb^{-1}$ at $\sqrt{s}=14\,\TeV$
if the signal efficiency is $>1/9$, 
while (as mentioned above) in the Type~III seesaw
there would not be enough events for discovery.

\section{Conclusions}
 Doubly charged Higgs bosons ($H^{\pm\pm}$),
which arise in the Higgs Triplet Model~(HTM) of neutrino mass generation,
are being searched for at the LHC\@.
 Separate searches have been carried out
for a pair of same-sign leptons ($\ell^\pm\ell^\pm$),
three leptons ($\ell^\pm\ell^\pm \ell^\mp$)
and  four leptons $(\ell^+\ell^+\ell^-\ell^-)$, 
all with comparable sensitivity to $m_{H^{\pm\pm}}^{}$.
 The neutral triplet scalars, $H^0$ and $A^0$,
are difficult to detect in the $7\,\TeV$ run of the LHC
for the case of degeneracy with $H^{\pm\pm}$.
 Positive values of a quartic coupling ($\lambda_4$) in the scalar potential
give rise to the mass hierarchy
$m_{A^0,H^0}^{} > m_{H^\pm}^{} > m_{H^{\pm\pm}}^{}$,
and in this scenario 
we showed that the branching ratios for the decays
$H^0\to H^\pm W^*$ and $A^0\to H^\pm W^*$
can be large ($\sim 100\%$ in the optimal case),
even for small mass splittings.
 The production processes
$q^\prime\overline{q}\to W^* \to H^\pm H^0$,
$q^\prime\overline{q}\to W^* \to H^\pm A^0$,
and $q\overline q\to Z^* \to H^0 A^0$,
together with the above decays
(and subsequently the decay $H^\pm\to H^{\pm\pm} W^*$)
would lead to production of $H^{++}H^{--}$
accompanied by several $W^*$ bosons.
 Of particular interest are the five-lepton (i.e.\ $W^*\to \ell\nu$)
and six-lepton channels (i.e.\ $W^*W^*\to \ell\nu\ell\nu$),
for which the background is known to be very small or negligible
for $\ell=e,\mu$.
 Importantly, 
the $6\ell$ channel can only arise from production mechanisms
which involve $H^0$ and $A^0$.
 For the LHC with $\sqrt s=7\,\TeV$,
and taking $m_{H^{\pm\pm}}^{}=300\,\GeV$,
we found maximum cross sections (including QCD $K$ factors)
of $5\,\fb$ and $0.8\,\fb$
for the $5\ell$ and $6\ell$ signatures respectively.
 Given the high detection efficiencies
in the ongoing searches for the $4\ell$ signature,
a signal in the $5\ell$ channel could be possible in the HTM\@.
 Consequently,
for the scenario of 
$m_{A^0,H^0}^{} > m_{H^\pm}^{} > m_{H^{\pm\pm}}^{}$,
all of the triplet scalars of the HTM could be discovered
in the $7\,\TeV$ run of the LHC in multi-lepton channels.
 For the LHC with $\sqrt s=14\,\TeV$
the maximum cross sections for the $5\ell$ and $6\ell$ channels are
$25\,\fb$ and $4\,\fb$ respectively,
which could allow a signal to be observed in the $6\ell$ channel.
 We will perform a detailed study of  the detection efficiencies
for the $5\ell$ and $6\ell$  signatures 
(as well as study other decay modes such as $W^*\to$ jets)
in a forthcoming work.

\section*{Acknowledgements}
 A.G.A was supported by a Marie Curie Incoming International Fellowship,
 FP7-PEOPLE-2009-IIF, Contract No.~252263.
SM is partially supported through the NExT Institute.
The work of H.S.\ was supported in part
by the Sasakawa Scientific Research Grant
from the Japan Science Society
and Grant-in-Aid for Young Scientists (B)
No.~23740210.

\end{document}